\documentclass[%
 aip,
 amsmath,amssymb,
 reprint,%
]{revtex4-1}

\usepackage{graphicx}
\usepackage{tabularx}
\usepackage{dcolumn}
\usepackage{bm}
\usepackage{float}
\usepackage[utf8]{inputenc}
\usepackage[T1]{fontenc}
\usepackage{mathptmx}
\usepackage{amsmath}
\usepackage{etoolbox}
\usepackage{xcolor}
\usepackage{xr}

\newcommand{\kon}{k_{\rm on}}
\newcommand{\koff}{k_{\rm off}}
\newcommand{\Non}{N_{\rm on}}
\newcommand{\Noff}{N_{\rm off}}

\makeatletter
\def\@email#1#2{%
 \endgroup
 \patchcmd{\titleblock@produce}
  {\frontmatter@RRAPformat}
  {\frontmatter@RRAPformat{\produce@RRAP{*#1\href{mailto:#2}{#2}}}\frontmatter@RRAPformat}
  {}{}
}%
\makeatother
\begin{document}

\preprint{AIP/123-QED}

\title[Noise in Nanofluidic Systems]{Decoding Noise in Nanofluidic Systems: \\ Adsorption versus Diffusion Signatures in Power Spectra}

\author{A. Drummond Young}
\affiliation{Physical and Theoretical Chemistry Laboratory, South Parks Rd, Oxford, OX1 3QZ UK}
\author{A. L. Thorneywork}
\email{alice.thorneywork@chem.ox.ac.uk}
\affiliation{Physical and Theoretical Chemistry Laboratory, South Parks Rd, Oxford, OX1 3QZ UK}
\author{S. Marbach}
\email{sophie.marbach@cnrs.fr}
\affiliation{CNRS, Sorbonne Université, Physicochimie des Electrolytes et Nanosystèmes Interfaciaux, F-75005 Paris, France}
\date{\today}

\begin{abstract}

Adsorption processes play a fundamental role in molecular transport through nanofluidic systems, but their signatures in measured signals are often hard to distinguish from other processes like diffusion. In this paper, we derive an expression for the power spectral density (PSD) of particle number fluctuations in a channel, accounting for diffusion and adsorption/desorption to a wall. Our model, validated by Brownian dynamics simulations, is set in a minimal but adaptable geometry, allowing us to eliminate the effects of specific geometries. We identify distinct signatures in the PSD as a function of frequency $f$, including a $1/f^{3/2}$ scaling related to diffusive entrance/exit effects, and a $1/f^2$ scaling associated with adsorption. These scalings appear in key predicted quantities -- the total number of particles in the channel and the number of adsorbed or unadsorbed particles -- and can dominate or combine in non-trivial ways depending on parameter values. Notably, when there is a separation of timescales between diffusion inside the channel and adsorption/desorption times, the PSD can exhibit two distinct corners with well-separated slopes in some of the predicted quantities. We provide a strategy to identify adsorption and diffusion mechanisms in the shape of the PSD of experimental systems on the nano- and micro-scale, such as ion channels, nanopores, and electrochemical sensors, potentially offering insights into noisy experimental data.

\end{abstract}

\maketitle

\section{Introduction}\label{sec:intro}

\begin{figure*}
    \centering
    \includegraphics[width=1\textwidth]{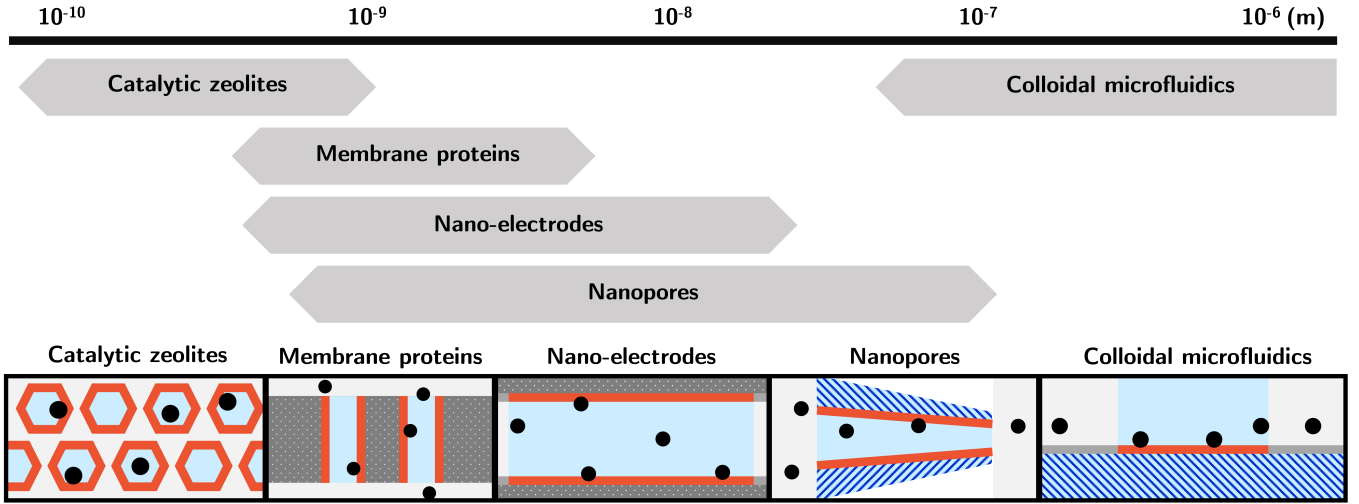}
    \caption{Adsorption and diffusion phenomena in nano- and micro- scale systems. In the schematics, particles (black) diffuse under confinement, with their relevant length scales highlighted. In these systems attractive interactions like adsorption (red) between the particles and the confining walls are also important, and add a degree of complexity.}
    \label{fig:lengthscales}
\end{figure*}

The diffusive passage of particles in confined geometries is a key phenomenon in many different systems -- see Fig.~\ref{fig:lengthscales}. A classic example is transport through protein channels, which facilitates the flow of ions and larger biomolecules across the cell membrane \cite{neher1976single, cooper1985theory, nikaido1992porins}. Inspired by this, protein channels have been used as a biotechnological tool for nanopore-based sequencing devices \cite{kasianowicz1996characterization, clarke2009continuous, bell2016digitally, deamer2016three}. Moreover, both biological and artificial nanopores have been increasingly developed for applications ranging from water filtration and blue energy harvesting \cite{werber2016materials, siria2017new} to functionalized single-molecule or single-particle sensing devices \cite{ying2022nanopore, dekker2007solid, goto2020solid, siwy2023nanopores, chazot2022optical}. A common feature of the measured signals in these systems is a relatively high level of noise. This arises, in part, because only small numbers of particles are involved at such small scales, but also arises due to the diffusive nature of particle motion \cite{kavokine2021fluids, verveen1974membrane}.

In many of the previous cases, attractive interactions, like adsorption between the transported particles and the confining surface itself play a key role (see red interfaces in Fig.~\ref{fig:lengthscales}). For example, protein channels make use of specific binding to encode selectivity \cite{luckey1980specificity, benz1987mechanism, andersen1995evaluation}, and in nanopore sensing technologies, adsorption has been used to prolong the dwell time of species in the channel, which aids detection \cite{howorka2001sequence, wei2012stochastic, yusko2011controlling, rotem2012protein, wiggin2008nonexponential, jetha2011long}. Adsorption can also transiently modify redox reactions for electrochemical detection \cite{singh2011stochastic}. In these cases, adsorption processes often act as an additional source of experimental noise. Adsorption to the pore walls may cause transient blockage of a nanopore and thus fluctuations in the measured ionic current \cite{andersen1995evaluation, nekolla1994noise}. It may also lead to fluctuations in the channel's surface charge, and hence in its ionic conductivity \cite{hoogerheide2009probing,secchi2016scaling}.

While detrimental from a sensing perspective, attempts have been made to use the link between characteristic noise signatures and molecular processes to reveal physical properties of the underlying system. For example, in nanopores, noise signatures could help identify the number of ionizable sites in a pore \cite{bezrukov1993current} or the presence of analyte or cation adsorption to the walls \cite{knowles2021current, singh2011stochastic,queralt2021specific, gravelle2019adsorption}. In nanoscale electrochemical devices, similar techniques have determined how reversible analyte adsorption to electrodes affects current noise \cite{katelhon2014noise, zevenbergen2009electrochemical}.

A methodology for understanding noisy signals common to many experimental systems is to investigate the power spectral density (PSD) of the measured signal. These investigations have revealed $1/f^\alpha$ scalings, where $\alpha \simeq 0.5 - 2$ and $f$ is the frequency, in biological and synthetic nanopores \cite{smeets2008noise, smeets2009low, knowles2019noise, fulinski1998non, fragasso20191, powell2009nonequilibrium, wohnsland19971, siwy2002origin} and electrochemical sensing systems \cite{zevenbergen2009electrochemical, katelhon2014noise}. Many theoretical approaches have revealed similar scaling laws, allowing us in principle to connect theories and experiments \cite{bezrukov2000particle, berezhkovskii2002effect, marbach2021intrinsic, gravelle2019adsorption, zevenbergen2009electrochemical, nestorovich2002designed, zorkot2016current, zorkot2016power}. Direct comparison is, however, limited to simple and specific cases. For instance, real experimental systems often exhibit heterogeneous transport properties at the single particle level \cite{knowles2024interpreting}, blurring the PSD signals. Often several noisy processes operate simultaneously, and models are used which require several fitting parameters, whose physical interpretation is limited \cite{robin2023disentangling}. Thus, in many cases theoretical approaches are still needed to help make concrete links between such scaling laws and molecular mechanism. 

In particular, it is difficult to rationalize the coupled effects of adsorption and diffusive transport. Previous theoretical work has investigated the effect on the PSD of a single binding site in a cylindrical channel \cite{berezhkovskii2002effect} and adsorption to the walls of a pore in combination with the effect of reservoirs \cite{gravelle2019adsorption, robin2023disentangling}. In these models, it can be difficult to distinguish the effects of adsorption from other model-specific features affecting particle dynamics, especially the geometry, which introduces several time- and length-scales \cite{marbach2021intrinsic, robin2023disentangling}. This has led some authors to linearly sum PSDs induced by uniquely diffusion or adsorption to account for experimental data \cite{zevenbergen2009electrochemical}, a likely flawed attempt in systems where the timescales for both phenomena can be similar \cite{knowles2021current, gadgil2004diffusion}. In particular, in experimental data involving adsorption and diffusion, it is still unclear when adsorption yields specific signatures in the PSD, as it is sometimes apparently linked to the emergence of two distinct slopes \cite{zevenbergen2009electrochemical, robin2023disentangling} and sometimes only one slope \cite{knowles2021current, secchi2016scaling}.

To answer this question, we introduce a simple framework to explore and separate the effects of diffusion and adsorption on power spectra. Our model involves non-interacting particles diffusing along a one-dimensional line, with a segment where particles can adsorb and become immobile -- see Fig.~\ref{fig:theorysetup}. Although the model is only an approximation to real physical systems, its simplicity is useful in two ways. Firstly, it allows us to isolate geometric effects from those of diffusion and adsorption within the channel, since in 1D the channel and reservoirs have only two length scales and therefore only two diffusive timescales. Secondly, it allows us to derive exact analytical expressions, validated by simulations, for the auto-correlation function and PSD of particle number fluctuations for all particles in the channel, as well as for free (unadsorbed) and bound (adsorbed) particles (Sec.~\ref{sec:theory}). The resulting PSDs are complex combinations of the purely diffusive and adsorbing processes, showing distinct scalings, like $f^{-3/2}$ and $f^{-2}$, which combine non-linearly. We rationalize the mechanisms behind each of these scaling laws: diffusive entrance/exit effects and adsorption/desorption phenomena, respectively. Crucially, when there is significant difference in timescales between diffusion across the channel and adsorption/desorption, the PSDs of some -- but not all -- of the predicted quantities exhibit two distinct slopes (Sec.~\ref{sec:results}). Finally, we discuss strategies to solve the inverse problem of determining adsorption properties from experimental data (Sec.~\ref{sec:conc}).

\section{Theoretical Model}\label{sec:theory}

\subsection{Setup to Probe the Effect of Adsorption}\label{sec:theorysetup}

\begin{figure}[H]
    \centering
    \includegraphics[width=0.48\textwidth]{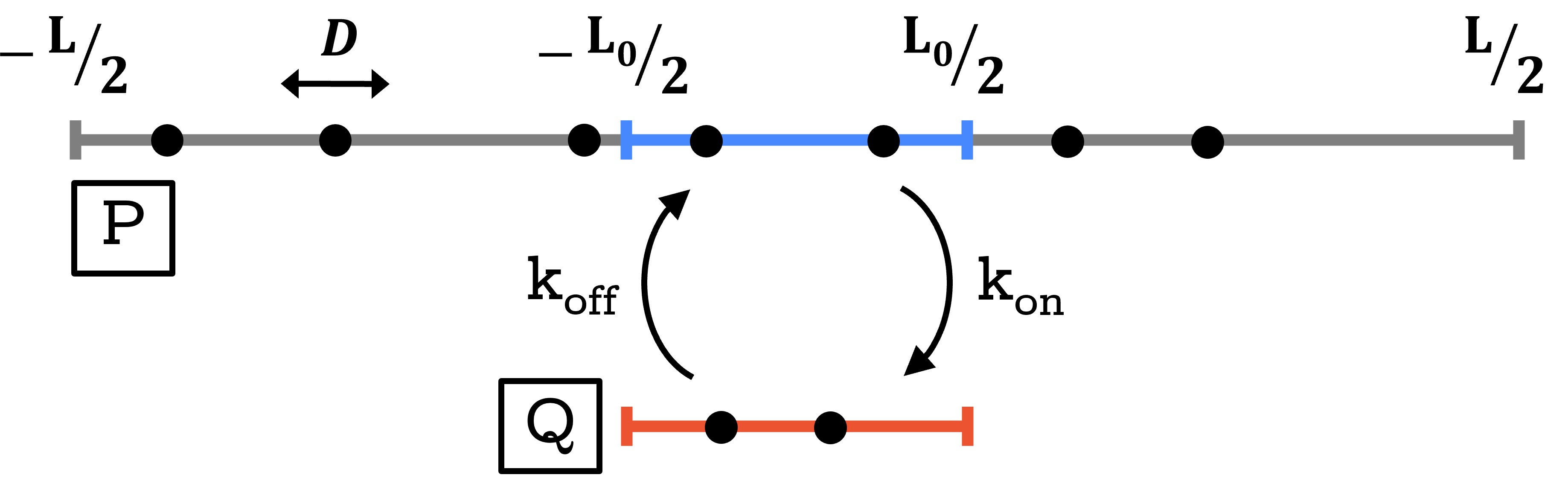}
    \caption{A representation of our model system of a channel connected to two reservoirs (line $P$) and adsorption on the channel wall (line $Q$). The channel has length $L_0$ while the total length of the system is $L$. Adsorption and desorption rates are set by $\kon$ and $\koff$, respectively.}
    \label{fig:theorysetup}
\end{figure}

We consider a 1D system of non-interacting particles diffusing with diffusion coefficient $D$ along a line $P$ with length $L$ -- see Fig.~\ref{fig:theorysetup}. In the center of the line, a region of length $L_0$ represents a ``channel region''. Inside the channel, in addition to diffusing, the particles can adsorb to a separate line $Q$ with rate $\kon$. On line $Q$, the particles cannot diffuse, but remain ``stuck''. Particles can unbind from $Q$ to line $P$ with rate $\koff$. Binding and unbinding rates are independent of the presence of other particles on either line. In this equilibrium scenario, the ratio of the rates corresponds to a Boltzmann factor, as $\kon/\koff \simeq e^{-e_0/k_B T}$ where $e_0$ is the effective adsorption energy and $k_B T$ the unit of energy. When adsorption is favorable, $e_0 \leq 0$, and $\kon \geq \koff$. The probability density of being in position $x$ at time $t$ on line $P$ is given by $p(x,t)$, and on line $Q$ by $q(x,t)$.

There are other possible approaches to include the effect of reservoirs in our system; for example, another line where particles can ``hop out'' of the system for $\lvert x \rvert >L_0/2$ (as in Ref.~[\onlinecite{marbach2021intrinsic}]). However, as the impact of the dimensions and geometry of reservoirs has been studied previously \cite{robin2023disentangling, gravelle2019adsorption} we focus on this 1D representation.

In choosing such a general 1D system rather than a specific geometry, we reduce the number of length scales in the system to two, $L$ and $L_0$. This reduces the number of corresponding diffusive timescales in the system to two, the minimum number possible in a system containing a channel and reservoirs. While more complicated 3D geometries may be more accurate representations of real systems, the additional length and timescales introduced would add complexities to the power spectra. Furthermore, even if the system is 3D, effective 1D behavior can still occur where particles are diffusing in a region with one length scale much shorter than the other(s) -- \textit{e.g.} in nanopores \cite{marbach2021intrinsic}. Even with only two length scales, we can change $L$ and $L_0$ at will to resemble systems where the adsorption region is large ($L\sim L_0$, as in electrochemical devices and assays) or small ($L\gg L_0$, as in many nanopore systems). Lastly, our 1D model allows us to solve for the predicted quantities exactly and gain physical insights.

Our goal is to establish the statistical properties of the number of particles inside the channel, $N(t)$, and how this number is split between free ($\Noff(t)$) and adsorbed ($\Non(t)$) particles. From the fluctuations of these numbers, we will calculate correlation functions and their associated power spectra. Each of these quantities -- freely diffusing particles, adsorbed particles, and their total -- may be of interest in different experimental settings, as different experimental methods may be sensitive to different particle subsets.

To make progress on analytics, we write the system of equations obeyed by $p(x,t)$ and $q(x,t)$ as:
\begin{equation}
\begin{cases}
    &\mathrm{for} \,\, |x| \leq L_0/2,\, \displaystyle\begin{cases} & \partial_t p = - k_{\rm on} p + k_{\rm off} q + D \partial_{xx}p \\
    &\partial_t q = + k_{\rm on} p - k_{\rm off} q
    \end{cases} \\
    &\mathrm{for} \,\, |x| > L_0/2,\, \partial_tp=D\partial_{xx}p.
    \label{eqn:setupequations}
\end{cases}
\end{equation}
The boundary conditions are \textit{(i)} continuity in $p$ at the edges of the channel, \textit{i.e.} at $x = \pm L_0/2$, and \textit{(ii)} no flux into the walls at the edges of the domain, \textit{i.e.} $\partial_x p = 0$ at $x = \pm L/2$. Note that the relevant continuous variable at the edges of the channel is $p$, rather than $p+q$, because there is no mechanism that allows particles to go directly from the reservoirs to being adsorbed inside the channel in one infinitesimal time step.

To quantify statistical properties, we need to establish the probabilities of transitioning from one state to another. For this reason, we start by considering a situation where a particle is initially located in the channel. It can begin either adsorbed, on line $Q$, or unadsorbed, on line $P$. Noting that the partition function of the system is $Z=L+L_0 (\kon/\koff)$, we can calculate the initial probability densities
\begin{equation}
\begin{aligned}
    p_0 &= p(|x| \leq L_0/2, t=0) = \frac{1}{Z} \\
    q_0 &= q(|x| \leq L_0/2, t=0) = \frac{\kon/\koff}{Z}
    \label{eqn:initialconditions}
\end{aligned}
\end{equation}
for these two cases, and $p(|x| \geq L_0/2, t=0) = 0$.

The system of equations in Eq.~\eqref{eqn:setupequations} can now be solved, working in Laplace space; full details of the calculation are in Appendix~\ref{app:rates}. From this solution we obtain expressions for $\hat{p}(x,s)$ and $\hat{q}(x,s)$, the Laplace transforms of variables $p(x,t)$ and $q(x,t)$, as
\begin{equation}
    \hat{p} = A_Ce^{m_1x}+B_Ce^{-m_1x}+\frac{\koff(p_0+q_0) + sp_0}{s(s+\koff+\kon)}
    \label{eqn:laplacep}
\end{equation}
and
\begin{equation}
\begin{aligned}
    \hat{q} = \frac{\kon}{s+\koff}(A_Ce^{m_1x}&+B_Ce^{-m_1x})\\ +&\frac{\kon(p_0+q_0) + sq_0}{s(s+\koff+\kon)}
    \label{eqn:laplaceq}
\end{aligned}
\end{equation}
where $s$ is the Laplace frequency, $m_0(s) = \sqrt{s/D}$ and $m_1(s) = \sqrt{s(s+\koff+\kon)/D(s+\koff)}$, and $A_C = B_C$ is a constant determined from the boundary conditions. Note that the expression for $\hat{p}$ is only true inside the channel where $|x| \leq L_0/2$.

In Sec.~\ref{sec:numberflucs}, we use these results to evaluate the total number fluctuations in the channel and their correlation function. Following this, in Sec.~\ref{sec:psd} we determine the power spectral density of the fluctuations.

\subsection{Number Fluctuations in the Channel}\label{sec:numberflucs}

In the following section, we calculate the fluctuating number of particles in the channel, $N(t)$, and then evaluate the corresponding correlation function. The same method can be used to calculate the correlation functions for the number of adsorbed particles $\Non(t)$ and the number of unadsorbed particles $\Noff(t)$; details are in Appendix~\ref{app:othercorrelations}.

We express the correlation function as
\begin{equation}
    C_N(t) = \langle N(t)N(0) \rangle = \mathcal{N} \psi_{C\rightarrow C}(t)
    \label{eqn:allparticlesprobs}
\end{equation}
where $\psi_{C\rightarrow C}(t)$ represents the probability that a particle starting in the channel is still in the channel at time $t$, and $\mathcal{N}$ is the total number of particles in the system. Note that the correct prefactor is indeed $\mathcal{N}$, the \textit{total} number of particles in the \textit{system} rather than the mean number in the channel. Formally, this originates from the fact that we are doing a canonical ensemble average here. Physically, any particle in the system has a probability to be in the channel at time $t=0$ and so we need to find the correlation function most generally for all of them -- and since particles are non-interacting, we can simply multiply the probability by the total number of particles inside the system.

Since we took initial conditions such that all particles started in the channel, we find $\psi_{C\rightarrow C}(t)$ simply by integrating the probability densities $p(x,t)$ and $q(x,t)$ over the channel length
\begin{equation}
    \psi_{C\rightarrow C}(t) =  \int_{-L_0/2}^{L_0/2} (p(x,t) + q(x,t)) dx.
    \label{eqn:psicc}
\end{equation}
We take the time derivative to obtain an expression for the correlation function as
\begin{equation}
\begin{aligned}
    \partial_t C_N(t) &= \mathcal{N} \int_{-L_0/2}^{L_0/2} \left(\partial_t p(x,t) + \partial_t q(x,t) \right)dx \\
    &= 2\mathcal{N} \int_{0}^{L_0/2} D \partial_{xx} p(x, t) dx \\
    &= 2\mathcal{N} D [\partial_x p(x, t)]^{L_0/2}_0
    \label{eqn:deltanall}
\end{aligned}
\end{equation}
and taking the Laplace transform we arrive at
\begin{equation}
    \hat{C}_N(s) = 2\mathcal{N} \frac{m_1}{m_0^2} A_C (e^{L_0 m_1/2} - e^{-L_0 m_1/2}).
    \label{eqn:correlationfuncfinal}
\end{equation}

\subsection{Spectral Density}\label{sec:psd}

The power spectrum $S_N(f)$ of number fluctuations at a given frequency $f$ is an important tool in analyzing their behavior at different timescales and is commonly used to characterize noisy experimental data in nanofluidic systems. It can be calculated from the correlation function of the number of particles in the channel using the Wiener-Khinchin theorem~\cite{alma991022195870107026}
\begin{equation}
    S_N(f) = \int_{-\infty}^{\infty} C_N(t) e^{-2 \pi i f t} dt.
    \label{eqn:wktheorem1}
\end{equation}
The correlation function for a stationary process like ours is even ($S_N(f) = 2\int_{0}^{\infty} C_N(t) e^{-2 \pi i f t} dt$), so using the Laplace transform
\begin{equation}
    S_N(f) = 4\Re[\hat{C_N}(2i\pi f)],
    \label{eqn:wktheorem2}
\end{equation}
with $\Re[\cdot]$ the real part of a scalar and where the additional factor of 2 comes from the fact that the power spectral density is calculated for positive frequencies only, $S_{>}(\omega)=2S(\omega)$, so that the total power is conserved \cite{alma991022195870107026}. Therefore, using the result in Eq.~\eqref{eqn:correlationfuncfinal} above, we find for the power spectrum of the number fluctuations in the channel
\begin{equation}
    S_N(f) = 8\Re \left[ \mathcal{N}\frac{m_1}{m_0^2} A_C \Big(e^{L_0m_1/2}-e^{-L_0m_1/2}\Big) \right].
\end{equation}
Substituting $A_C$, and applying the same transformation to obtain the power spectrum of the number of bound or free particles in the channel, we obtain
\par\vspace{10pt}
\begin{widetext}
\begin{equation}
\begin{gathered}
    S_N(f) = 8\Re\Bigg[\frac{\mathcal{N} m_1 (\koff(p_0+q_0)+s p_0)}{s m_0 (s+\koff+\kon)} \frac{(e^{L m_0}-e^{L_0 m_0})(1-e^{L_0 m_1})}{(m_0-m_1)(e^{L m_0}-e^{L_0(m_0+m_1)})+(m_0+m_1)(e^{L m_0 + L_0 m_1}-e^{L_0 m_0})}\Bigg],
 \\
    S_{\rm on}(f) = 8\Re\Bigg[ \frac{\mathcal{N} \koff q_0}{s(s+\koff+\kon)} \Bigg( -\frac{L_0}{2} +\frac{m_0 \kon}{m_1 (s+\koff)} \frac{(e^{L m_0}-e^{L_0 m_0})(1-e^{L_0 m_1})}{(m_0-m_1)(e^{L m_0}-e^{L_0(m_0+m_1)})+(m_0+m_1)(e^{L m_0 + L_0 m_1}-e^{L_0 m_0})} \Bigg) \Bigg],
 \\
     S_{\rm off}(f) = 8\Re\Bigg[ \frac{\mathcal{N} \kon p_0}{s(s+\koff+\kon)} \Bigg( -\frac{L_0}{2} +\frac{m_0 (s+\koff)}{m_1 \kon} \frac{(e^{L m_0}-e^{L_0 m_0})(1-e^{L_0 m_1})}{(m_0-m_1)(e^{L m_0}-e^{L_0(m_0+m_1)})+(m_0+m_1)(e^{L m_0 + L_0 m_1}-e^{L_0 m_0})} \Bigg) \Bigg].
    \label{eqn:psd}
\end{gathered}
\end{equation}
\end{widetext}
Note that $s = 2i\pi f$ in the above expression. Eq.~\eqref{eqn:psd} is the main theoretical result of this paper. In its current form, it is hardly amenable to physical interpretation, however, and our goal in the following sections is to uncover limiting regimes and mechanistic behavior from this expression.

\section{Results}\label{sec:results}

Our system is governed by four key timescales:
\begin{itemize}
    \item Time to diffuse across the system: $\tau_{\rm sys}=L^2/2D$
    \item Time to diffuse across the channel: $\tau_{\rm cha}=L_0^2/2D$
    \item Adsorption timescale: $\tau_{\rm ads}=1/\kon$
    \item Desorption timescale: $\tau_{\rm des}=1/\koff$
\end{itemize}
These timescales can be estimated for many nano-technological systems and we report orders of magnitude in Table~\ref{tab:param_vals}. Remarkably, there is no standard case, and relative magnitudes can vary strongly from one system to another. Our goal is not to imitate these systems exactly with our model, but rather to illustrate the diversity of behavior which can be observed in power spectra with only two physical processes.
\begin{table}
    \centering
    \renewcommand{\arraystretch}{1.2}
    \setlength{\tabcolsep}{5pt}
    \begin{tabular}{>{\raggedright\arraybackslash}m{0.5\linewidth}ccc}
        \hline
         \textbf{Example} & \textbf{\boldmath$\tau_{\rm cha}$ (s)} & \textbf{\boldmath$\tau_{\rm ads}$ (s)} & \textbf{\boldmath$\tau_{\rm des}$ (s)}\\
         \hline 
         \multicolumn{4}{c}{Case 1: $\tau_{\rm cha} \gg \tau_{\rm ads} \sim \tau_{\rm des}$} \\
         \hline
         Molecules reacting in nanofluidic cavities \cite{zevenbergen2009electrochemical} & $10^{0}$ & $10^{-1}$ & $10^{-3}$\\
         \hline
         \multicolumn{4}{c}{Case 2: $\tau_{\rm cha} \ll \tau_{\rm ads} \sim \tau_{\rm des}$} \\
         \hline
         Sugar transport in transmembrane proteins \cite{nekolla1994noise, kullman2002transport} & $10^{-8}$ & $10^{-2}$ & $10^{-3}$\\
         Adapter adsorption inside nanopores \cite{gu2001prolonged} & $10^{-5}$ & $10^{-2}$ & $10^{-3}$\\
         Nano-electrochemical devices \cite{katelhon2014noise} & $10^{-7}$ & $10^{-4}$ & $10^{-5}$\\
         \hline
         \multicolumn{4}{c}{Case 3: $\tau_{\rm cha} \sim \tau_{\rm ads} \sim \tau_{\rm des}$} \\
         \hline
         PEG adsorption in glass nanopores \cite{knowles2021current} & $10^{-4}$ & $10^{-4}$ & $10^{-2}$\\
         DNA hybridization in microarrays \cite{gadgil2004diffusion} & $10^{3}$ & $10^{2}$ & $10^{4}$\\
         \hline
         \multicolumn{4}{c}{Case 4: $\tau_{\rm ads} \gg \tau_{\rm cha} \sim \tau_{\rm des}$} \\
         \hline
         Biomolecule sensing in porous silicon membranes \cite{zhao2016flow} & $10^{0}$ & $10^{-1}$ & $10^{6}$\\
         Gas phase catalysis in zeolites \cite{hansen2009analysis} & $10^{-2}$ & $10^{-2}$ & $10^{2}$\\
        \hline
    \end{tabular}
    \caption{Typical values for the timescales of channel diffusion, adsorption, and desorption, in a number of different experimental systems. We do not include $\tau_{\rm sys}$ here; it is harder to determine in general and in most cases will either be similar to the channel length ($L \sim L_0$) or much greater ($L \gg L_0$, in which case the limit of an infinite system is approached).}
    \label{tab:param_vals}
\end{table}
In the following, we start by investigating limit cases of pure diffusion or only adsorption. While the scaling laws associated with these processes in isolation are known, the full problem of interest here involves a non-trivial combination of these two canonical mechanisms. As such, we provide a detailed  explanation of how scalings emerge from fluctuations associated with these separate phenomena as context for later results. Scaling laws for adsorption and diffusion are also accessible experimentally and provide a useful reference point for experimentalists.

Related to this, we work in the following reduced units. In all cases the characteristic length scale is $\ell_0 = L_0$, and the characteristic timescale is $\tau_0 = \tau_{\rm cha} = L_0^2/2D$, except in the case of pure adsorption/desorption where $D=0$, in which case we use $\tau_0 = 1/f_0$, where $f_0 = (\kon+\koff)/2\pi$ and represents the corner frequency in the PSD. We compare the analytical results, derived above, to particle-based simulations of the system (see Appendix~\ref{app:methods} for details on computational methods). 
Again, we generally choose a 1D geometry for our simulations to provide a robust verification of our analytical results.

\subsection{Pure Diffusion}\label{sec:pure_diffusion}

We first verify that our results recover the expected behavior for the case of pure diffusion of particles in a channel \cite{bezrukov2000particle, marbach2021intrinsic}. This scenario corresponds to setting $\kon=0$ and $q_0=0$. Here, we obtain $S_{\rm on}(f) = 0$, $S_{\rm off}(f) = S_N(f)$ and
\begin{equation}
    S_N(f) = 4\Re\Bigg[ \frac{\mathcal{N} p_0}{m_0 s} \frac{(e^{Lm_0}-e^{L_0m_0})(e^{-L_0m_0}-1)}{(e^{Lm_0}-1)} \Bigg]
    \label{eqn:diffusionlimit}
\end{equation}
where we recall that in the expression above $s = 2i \pi f$.

\begin{figure}
    \centering
    \includegraphics[width=1\linewidth]{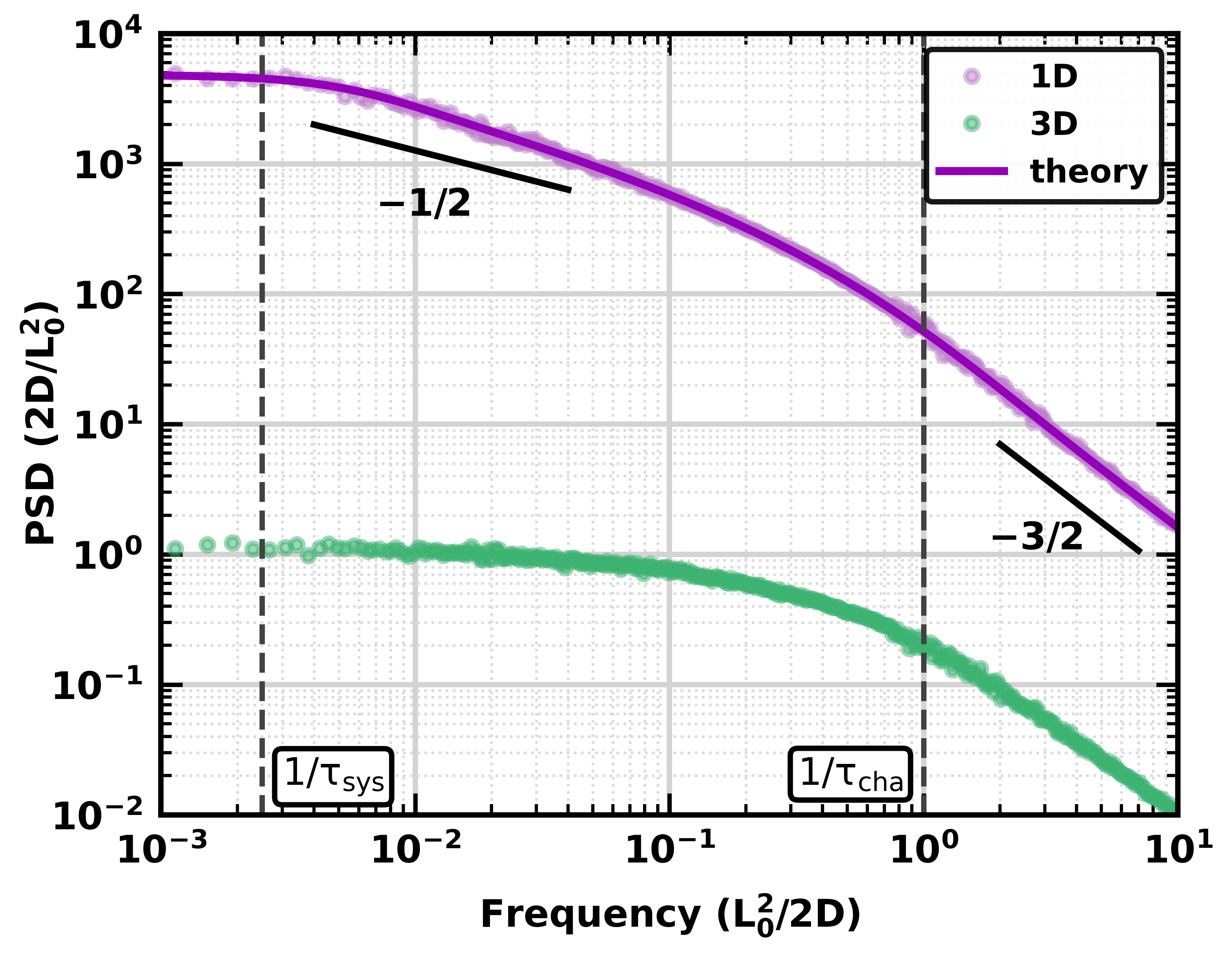}
    \caption{Power spectral density of number fluctuations for the purely diffusive case, with $L=20\ell_0$. Dashed lines indicate $\tau_{\rm sys}=L^2/2D$ and $\tau_{\rm cha}=L_0^2/2D$. Expected scalings for high and intermediate frequency regimes of $f^{-3/2}$ and $f^{-1/2}$ respectively are indicated. Data in purple shows the spectrum for the total number of particles from simulations. Data in green shows the corresponding spectrum for simulations in 3D, with the same parameters. The 3D result plateaus for frequencies smaller than $1/\tau_{\rm cha}=2D/L_0^2$.}
    \label{fig:intermed_scaling}
\end{figure}

The theoretical (line) and simulation (dots) spectra for the total number of particles are shown in Fig.~\ref{fig:intermed_scaling} (purple) and are in perfect agreement. The spectrum clearly involves multiple regimes; more specifically, a plateau at low frequencies and then two scaling regimes with exponents $-1/2$ and $-3/2$. The scaling regimes can be interrogated by considering the limits of Eq.~\eqref{eqn:diffusionlimit}.

Firstly, as $f\rightarrow0$ we find:
\begin{equation}
\begin{aligned}
    S_N(f) &= \frac{\mathcal{N} p_0 (L-L_0)^2 L_0^2}{3 D L}
    \label{eqn:diffusionplateau}
\end{aligned}
\end{equation}
\textit{i.e.} there is a plateau at low frequencies.
At high frequencies as $f\rightarrow \infty$, we find instead that the PSD tends to
\begin{equation}
\begin{aligned}
    S_N(f) &= \frac{\mathcal{N} p_0 \sqrt{D}}{(\pi)^{3/2}} \frac{1}{f^{3/2}}
\end{aligned}
\end{equation}
with a scaling as $f^{-3/2}$. Between these two regimes, the slope of the PSD appears to vary steadily between the two limits; however, this masks a well-defined underlying scaling. To find this, we take $L\rightarrow \infty$ and $f\rightarrow0$
\begin{equation}
\begin{aligned}
    S_N(f) &= \frac{\mathcal{N} p_0 L_0^2}{\sqrt{D \pi}} \frac{1}{\sqrt{f}}
\end{aligned}
\end{equation}
predicting a scaling as $f^{-1/2}$. Physically, we can interpret these different regimes as follows.

\paragraph{Plateau is linked to the mean correlation time.} For low frequencies, corresponding to timescales longer than the time for a particle to diffuse across the system $\tau_{\rm sys}$, we see a plateau (white noise). This is due to the finite size of the system -- fluctuations in particle number are eventually uncorrelated because the particles have had sufficient time to equilibrate outside of the channel by diffusion. The plateau value of a PSD spectrum corresponds to a characteristic timescale as $S_N(f) = \mathcal{N} \tau_{\rm corr}$ where $\tau_{\rm corr}$ is the mean correlation time of the number fluctuations, 
\begin{equation}
    \tau_{\rm corr} = \int_0^{\infty} \frac{C_N(t)}{\mathcal{N}} \mathrm{d}t.
\end{equation}
Remembering that $p_0 = 1/L$ and taking $L \gg L_0$ for simplicity, we find $\tau_{\rm corr} = L_0^2/3D = 2\tau_{\rm cha}/3$. The plateau amplitude thus depends on the mean time particles spend inside the channel. It is interesting that the amplitude does not depend on system size $L$ but actually probes the channel size $L_0$, and the only dependence of the plateau on $L$ is the frequency at which it begins.

\paragraph{High frequency $f^{-3/2}$ is linked to diffusive particle crossings in and out of the channel.} At high frequencies, corresponding to timescales shorter than the time for a particle to diffuse across the channel, $\tau_{\rm cha}$, we see a characteristic $f^{-3/2}$ scaling \cite{marbach2021intrinsic, LAX1960248, zevenbergen2009electrochemical, berezhkovskii2002effect, minh2023ionic}. Its origin is particles crossing from one region to another (\textit{i.e.} from channel to reservoir and \textit{vice versa}), by diffusion. These events correspond to short time scales and are essentially independent of the system geometry.

\paragraph{Intermediate $f^{-1/2}$ is linked to re-entrance effects.} The intermediate regime between these two timescales -- where we observe a slope in the PSD lower than $3/2$, but have not yet reached a plateau -- is a signature of a $f^{-1/2}$ scaling \cite{marbach2021intrinsic, MacFarlane1950807, Burgess1953334, VANVLIET1958415, LAX1960248, vossflicker1976}. This scaling arises from re-entrance effects, \textit{i.e.} it emerges directly from the fact that particles in 1D always return to the origin. This intermediate scaling depends on dimensionality. In an $n$-dimensional system, the probability of particle returns goes as $1/t^{n/2}$, so in 1D the spectral scaling is $f^{-1/2}$, in 2D it is $\rm{ln}(1/\textit{f})$, and in 3D it is $1/f^{0}$, \textit{i.e.} a plateau \cite{vossflicker1976, MacFarlane1950807, Burgess1953334}. The cross-over from 1D to 3D is the point at which particles no longer inevitably return to their origin, so the re-entrance correlations disappear, as shown in Fig.~\ref{fig:intermed_scaling} (green dots). In more realistic designs, and even in 3D systems, there may be effective 1D or 2D scaling behavior if one or more dimension is small relative to the others \cite{vossflicker1976}. In our 1D model, a plateau below  $1/\tau_{\rm sys}$ appears due to finite system size. These timescales may be inaccessible in experimental systems. Overall, these low frequency scalings are linked to the reservoir geometry, which is not our focus here.

\subsection{Only Adsorption/Desorption}

The second key source of fluctuations in the full problem is the adsorption/desorption process itself. To consider characteristics of this in isolation, \textit{i.e.} adsorption in a system with no diffusion, we set $D=0$ in the original system of equations Eq.~\eqref{eqn:setupequations} and solve the system for $p$ and $q$. Here, there will naturally be no fluctuations in the total number of particles in the channel region ($S_N(f)=0$), but there are still fluctuations in the number of bound and free particles. As both quantities are directly related to one another in that case, their associated spectra are equal, $S_{\rm on}(f) = S_{\rm off}(f)$. Additionally, we can use $p_0$ and $q_0$ interchangeably as they are related by $p_0\kon=q_0\koff$. In this situation, we obtain
\begin{equation}
    S_{\rm on/off}(f) = \frac{\mathcal{N}_{\rm off} k_{\rm on}}{\pi^2} \frac{1}{f^2 + f_0^2}
    \label{eqn:adsdeslimit}
\end{equation}
where $\mathcal{N}_{\rm off} = \mathcal{N} L_0 p_0$ is the mean number of free particles, and we recall that $f_0 = (\koff+\kon)/2\pi$. This is consistent with works which investigate exclusively adsorption/desorption processes \cite{zevenbergen2009electrochemical, nestorovich2002designed}.

\begin{figure}
    \centering
    \includegraphics[width=1\linewidth]{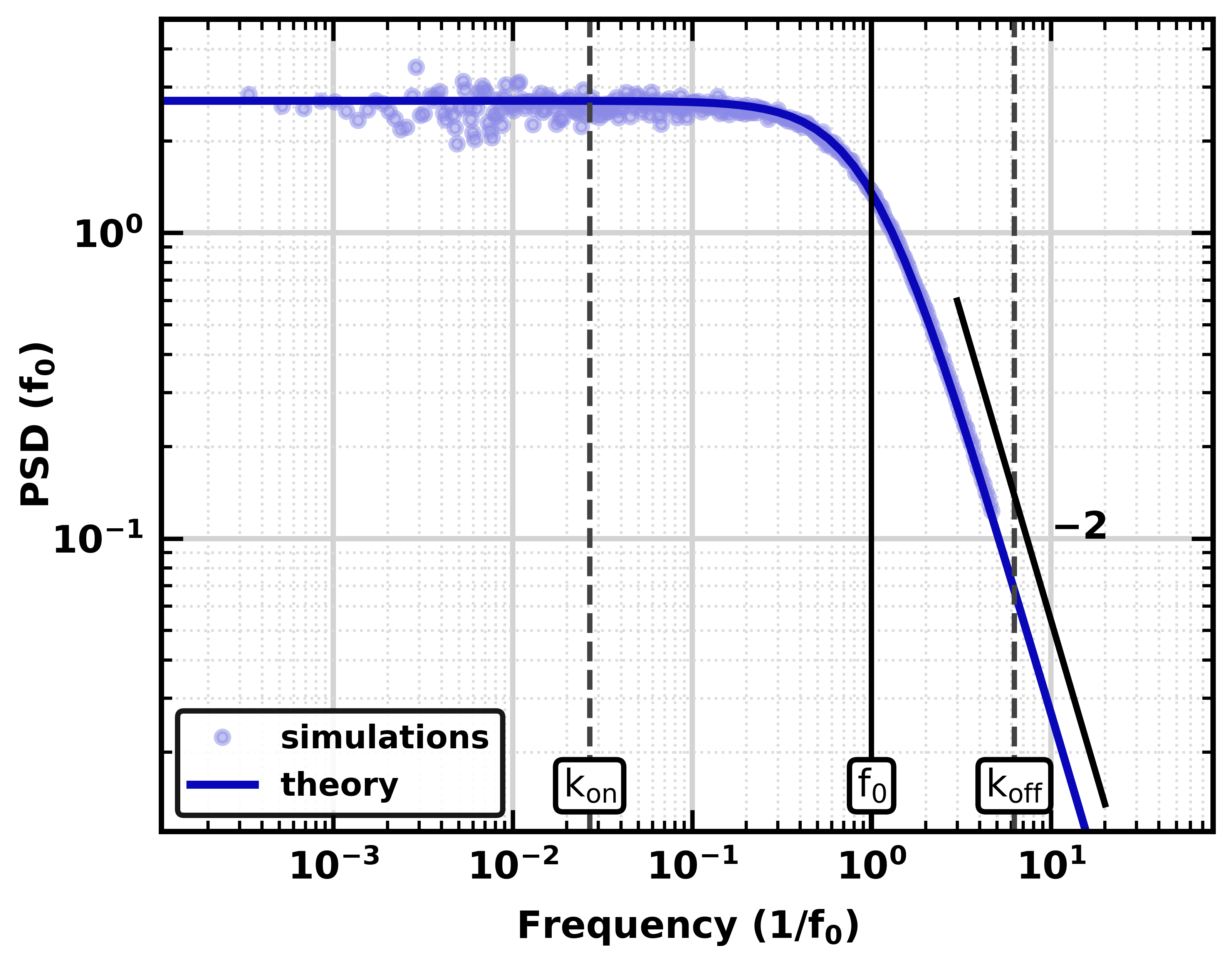}
    \caption{Power spectral density of number fluctuations with no diffusion and only adsorption/desorption processes. Spectrum for fluctuations in the number of unadsorbed particles is shown, which is equivalent to that for adsorbed particles. Dashed lines indicate $\tau_{\rm ads}=1/\kon$ and $\tau_{\rm des}=1/\koff$. A solid line indicates $f_0$, the corner frequency, which is the mean of the two rates $f_0=(\kon+\koff)/2\pi$. There is a clear $1/f^2$ scaling at high frequencies and a plateau below $f_0$. In this case, $L=L_0=\ell_0$, $\kon=0.027f_0$, and $\koff=6.3f_0$. We use separated adsorption and desorption rates to demonstrate the dependence of the corner primarily on the faster rate (here, $\koff$).}
    \label{fig:adsdes}
\end{figure}

Spectra obtained from Eq.~\eqref{eqn:adsdeslimit} and simulations for the pure adsorption/desorption scenario are compared in Fig.~\ref{fig:adsdes}.
The shape of the spectrum is a Lorentzian, scaling as $1/f^2$ at high frequencies. This can be intuitively understood as, at very short timescales, the fluctuating signal appears as individual step functions. Transforming into frequency space, a step function becomes $1/f$, which is then squared to obtain the PSD and $1/f^2$ scaling. At low frequencies, or long timescales, the fluctuating signal is ``smeared out'' and resembles a white noise, or a plateau in the PSD. The binding/unbinding process corresponds to a \textit{single} shot noise process, which again intuitively makes sense because binding and unbinding events are uncorrelated and thus follow a Poisson distribution.

The cross-over frequency between the two regimes is $f_0 = (\kon+\koff)/2\pi$, \textit{i.e.} the mean of the adsorption and desorption rates. This can be understood by considering the situation where one of the two processes becomes very fast. For \textit{e.g.} fast desorption $\koff \gg \kon$, the probability of an adsorption event being followed later in time by another adsorption event is independent of the lifetime of the adsorbed state, becoming a pure shot noise process in the limit of zero adsorption time. As such, correlations in the noise only persist for frequencies larger than the faster rate. In contrast, as lifetimes on and off the surface become comparable, the probability of adsorption will be influenced by previous adsorption events, resulting in a corner frequency that is the average of the two rates.
The plateau value again corresponds to a mean correlation time, as $S_{\rm on/off}(f\rightarrow 0) = \mathcal{N}_{\rm off} q_0 \tau_0$, where the relevant correlation time is $\tau_0=1/(2\pi f_0) = 1/(k_{\rm on} + k_{\rm off})$, the mixture between the two timescales.

\begin{figure*}
    \centering
    \includegraphics[width=1\textwidth]{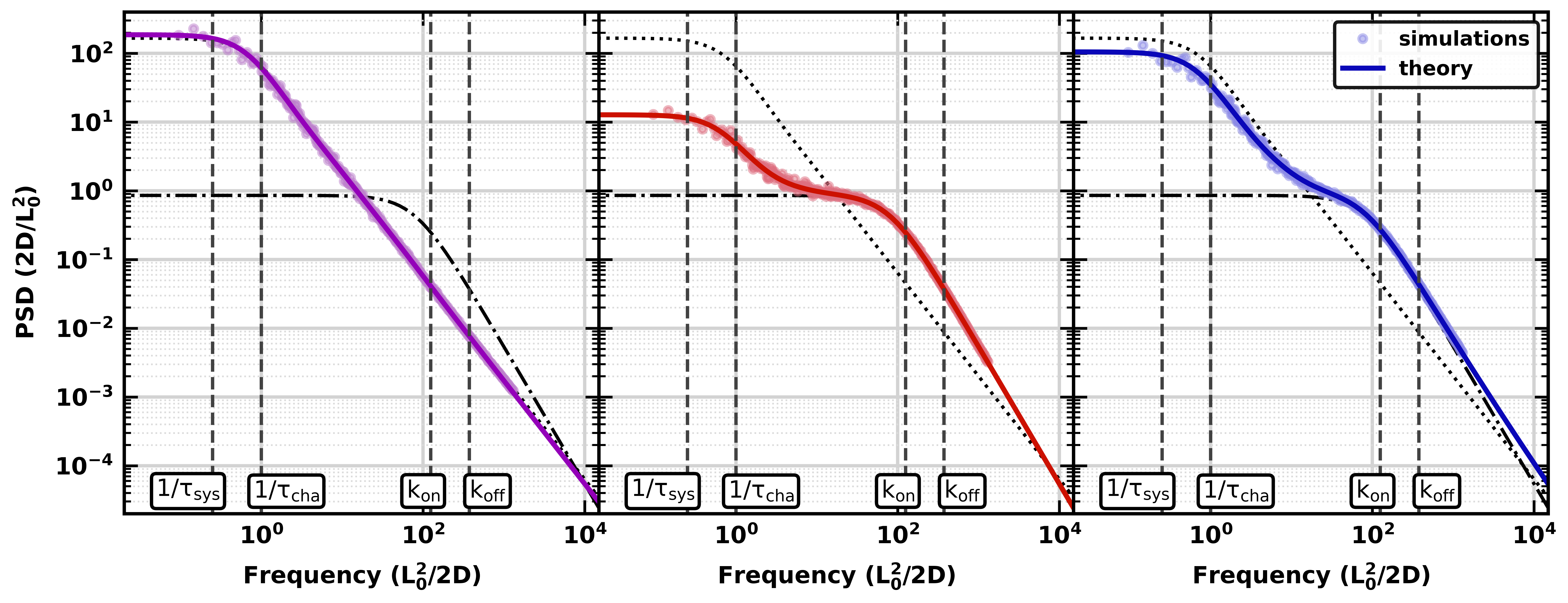}
    \caption{\textbf{Slow diffusion, fast adsorption/desorption: Case 1 -- $\tau_{\rm cha} \gg \tau_{\rm ads} \sim \tau_{\rm des}$.} Left: total PSD. Middle: bound PSD. Right: free PSD. We use $L=2\ell_0$, $\kon=125/\tau_0$, and $\koff=375/\tau_0$. Note here that $\kon \leq \koff$. The dotted curve indicates the pure diffusive case with the same parameters (but $\kon=0$). Similarly, the dash-dotted curve indicates the pure adsorption/desorption case with the same parameters (but $D=0$). Relevant timescales are labeled and marked with dashed vertical lines.}
    \label{fig:case_1a}
\end{figure*}

Similar noise processes and therefore spectral signatures are observed in various other physical systems. For example, this noise is similar to random telegraph or ``popcorn'' noise -- noise observed in semiconductors where there are sudden transitions between two (or sometimes more) current levels \cite{machlup1954noise}. Analogously, here, this noise is due to particles having only two states, ``on'' or adsorbed, and ``off'' or unadsorbed. Equally, the process can be mapped to scenarios where noise arises from signals involving ``free particle pulses''.  Here, the ``pulse'' is  the mean number of free particles $\mathcal{N}_{\rm off}$ and occurs with mean frequency $k_{\rm on}$, which corresponds to the mean time particles stay free. Considering that $\mathcal{N}_{\rm off} k_{\rm on} = \mathcal{N}_{\rm on} k_{\rm off}$, the process can also be interpreted as ``bound particle pulses''. The subtlety is that the amplitude of each of these pulses decays exponentially with a timescale which is the mixture of the two timescales at play $\tau_0$. The exponential distribution of the pulses is a specificity of this system, and thus, compared to other processes whose length in time are constant, as in Ref.~[\onlinecite{knowles2024interpreting}], the shape of the PSD has no oscillations and is a true Lorentzian \cite{verveen1974membrane}.

\subsection{Full Case}\label{sec:fullcase}

We now consider the full problem where both diffusion and adsorption processes lead to fluctuations. Notably, in this case the results from the previous two sections do not add linearly, \textit{i.e.} we have $S_{N}(f) \neq S_{\rm on}(f) + S_{\rm off}(f)$. In practice, this is due to the presence of cross-correlations in the fluctuations, and the spectrum of the total number of particles in the channel is the sum of 4 terms
\begin{equation}
    S_{N}(f) = S_{\rm on}(f) + S_{\rm off}(f) + S_{\rm{on \rightarrow off}}(f) + S_{\rm{off \rightarrow on}}(f).
    \label{eqn:additivity}
\end{equation}
The cross-correlation terms are in general non-trivial (see Appendix~\ref{app:othercorrelations}). Although we now observe a much more complicated picture, both in the analytical expressions in Eq.~\eqref{eqn:psd} and in the shapes and behavior of the PSDs, some distinguishable ``common'' scaling regimes are reached which we discuss below. To build intuition, it is useful to consider the following limiting cases:
\begin{enumerate}
    \item Slow diffusion, fast adsorption/desorption: $\tau_{\rm sys} \sim \tau_{\rm cha} \gg \tau_{\rm ads} \sim \tau_{\rm des}$
    \item Fast diffusion, slow adsorption/desorption: $\tau_{\rm sys} \sim \tau_{\rm cha} \ll \tau_{\rm ads} \sim \tau_{\rm des}$
    \item All processes have comparable timescales:
    $\tau_{\rm sys} \sim \tau_{\rm cha} \sim \tau_{\rm ads} \sim \tau_{\rm des}$
    \item Fast adsorption compared to desorption: $\tau_{\rm des} \gg \tau_{\rm sys} \sim \tau_{\rm cha} \gg \tau_{\rm ads}$
\end{enumerate}
We refer to the power spectra of the total particles in the channel, the adsorbed particles, and the free particles in the channel as the ``total PSD'', ``bound PSD'', and ``free PSD'', respectively. The first two cases we consider in detail in the main text, and the second two in Appendix~\ref{app:extraresults}. Note that in the following, the dimensionless timescale $\tau_0 = L_0^2/2D$ refers to the time to diffuse across the channel.

\begin{figure*}
    \centering
    \includegraphics[width=1\textwidth]{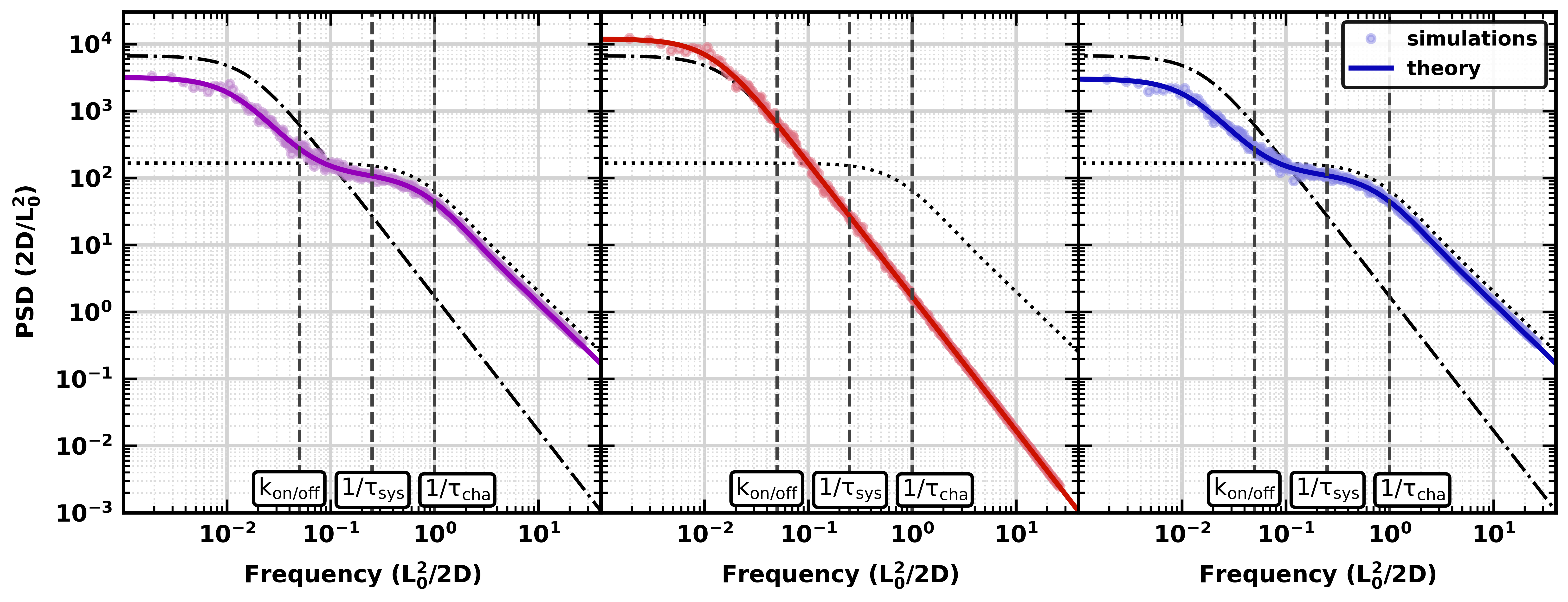}
    \caption{\textbf{Fast diffusion, slow adsorption/desorption: Case 2 -- $\tau_{\rm cha} \ll \tau_{\rm ads} \sim \tau_{\rm des}$ } Left: total PSD. Middle: bound PSD. Right: free PSD. We use $L=2\ell_0$, $\kon=\koff=0.05/\tau_0$. The color/line code is the same as that of Fig.~\ref{fig:case_1a}.}
    \label{fig:case_2}
\end{figure*}

\subsubsection{Diffusion Slower than Adsorption/Desorption}\label{sec:resultsDSAD}

We first explore the case where diffusion is slower than both adsorption and desorption processes. Physically, this could manifest as the adsorption/desorption of redox-active species in nanofluidic electronic devices, which can be fast with respect to their diffusion \cite{zevenbergen2009electrochemical}.

Results (both theory and simulations) are shown in Fig.~\ref{fig:case_1a}, where the corresponding curves for a pure diffusion (dotted black line) or adsorption (dash-dotted black line) process are shown for comparison. The total PSD is close to the purely diffusive case, with a corner frequency around the diffusion timescale and a high frequency scaling of $f^{-3/2}$. It is perhaps counter-intuitive that the total PSD, $S_{N}$, does not show the corner frequency linked to adsorption given that the pure adsorption curve has a higher magnitude than the diffusion curve at high frequencies. Mathematically, this is a result of cross-correlations, which may be negative, \textit{i.e.} the power spectra are not additive, $S_{N} \neq S_{\rm on} + S_{\rm off}$ (see Appendix~\ref{app:othercorrelations}). Physically, the reason for this is that adsorption and desorption processes are so fast that they mostly affect the relative distribution of bound or free particles, but these processes have ``averaged'' out before a particle may escape the channel by diffusion and contribute fluctuations to the total PSD.

Bound and free PSDs are similar to each other and both display two corner frequencies, one around the diffusive timescales and one at the binding timescales. At low frequencies, both PSDs plateau at values which sum to less than that of the purely diffusive case, again highlighting non-additivity. In the intermediate regime between the two corners, there is a transient scaling regime; this appears as $f^{-3/2}$ due to diffusion for the free PSD, but is sufficiently short that this may not always be obvious in experimental systems. At high frequencies, the bound PSD is dominated by a $f^{-2}$ scaling, which intuitively makes sense as the faster processes of adsorption and desorption dominate on short timescales.

There is an important distinction between the bound and free PSDs; at high frequencies, although the free PSD follows a $f^{-2}$ scaling just above the adsorption/desorption rates, it eventually returns to a $f^{-3/2}$ scaling. This is not obvious over a short range of frequencies, as the deflection is very slow, and occurs above $f\sim L_0^2k_{\rm on}^2/D \pi\sim \tau_{\rm cha}/\tau_{\rm ads}^2$. In all cases, in fact, the free PSD returns to a $f^{-3/2}$ scaling at high frequencies as a limiting case. This is due to the fact that if particles are ``close'' enough to the channel boundary, they will diffusively enter/exit the channel at infinitely fast timescales, and hence a diffusive $f^{-3/2}$ scaling has to be recovered at high enough frequencies. Crucially, the slow transition between these scaling regimes would not be clear in an experiment which can only probe a few decades of frequency space, so it is possible that the $f^{-3/2}$ scaling is missed entirely or mistaken for having an exponent between $-1.5$ and $-2.0$. The absence of a sharp corner also masks this underlying shift in scalings.

The bound PSD is very similar to the spectrum of fluctuations in the current measured in Ref.~[\onlinecite{zevenbergen2009electrochemical}]. In that nanofluidic system, it is likely that the relevant quantity being probed by the current is the number of species adsorbed onto the electrodes, which means our results for the bound PSD are in good agreement. Our model rationalizes these results in the context of other fluctuating quantities in the system. Finally, we note that the ordering of adsorption and desorption timescales does not significantly affect the shape of the power spectra; the complementary case where adsorption is faster than desorption ($\kon \gg \koff$) is discussed in Appendix~\ref{app:extraresults}, in particular in Fig.~\ref{fig:case_1b}.

\subsubsection{Diffusion Faster than Adsorption/Desorption}\label{sec:resultsDFAD}

\begin{figure*}
    \centering
    \includegraphics[width=1\textwidth]{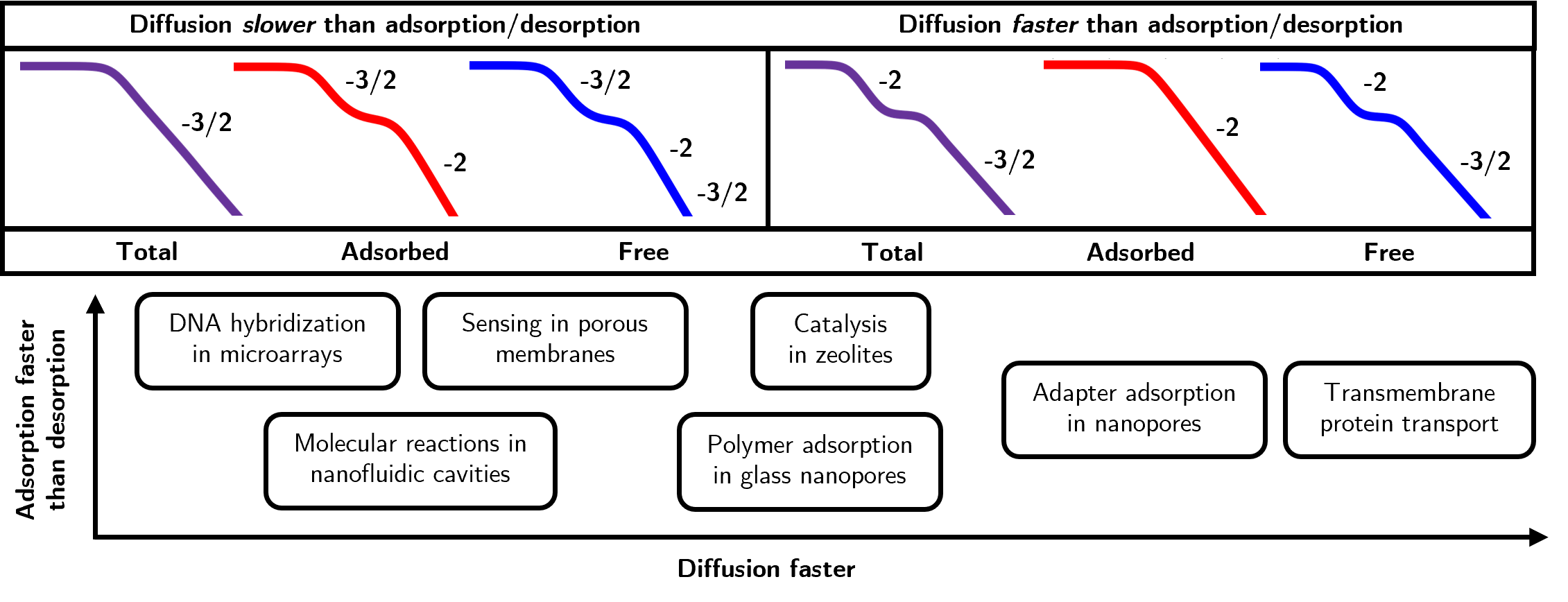}
    \caption{(Top) Characteristic shapes and scalings of the PSDs identified in this work. (Bottom) Corresponding physical systems in different diffusion/adsorption regimes.}
    \label{fig:phase_diagram}
\end{figure*}

We now turn to the situation where diffusion is faster than adsorption/desorption. This case is the one considered in Ref.~[\onlinecite{berezhkovskii2002effect}]. Physically, this ordering of timescales is common in nanopores, both biological and engineered. For example, it occurs naturally in the transport of sugars through maltoporins (LamB transmembrane proteins) \cite{kullman2002transport, nekolla1994noise} and also in the use of adapters, such as beta cyclodextrin, in nanopore sensors for detection of organic molecules \cite{gu2001prolonged}. Nanoscale electrical devices where analytes reversibly bind to the electrodes may also involve fast diffusion relative to the adsorption timescales \cite{katelhon2014noise}.

Representative spectra are shown in Fig.~\ref{fig:case_2}. Here, we only consider the case where $\kon=\koff$, because varying $\kon$ and $\koff$ does not significantly affect the shape or behavior of the PSDs.

We observe that the total PSD is more sensitive to the adsorption/desorption process compared to Case 1 where diffusion was slower than adsorption/desorption. In particular, it now shows two corners and two slopes, with the low-frequency corner determined by the adsorption/desorption timescale and high frequency corner by the diffusive timescale(s). Again, there is a transient intermediate scaling regime between the two corners.

In this case, the total and free PSDs have similar shapes whereas previously the free and bound PSDs were similar. Here, with fast diffusion, the number of bound particles is less dependent on the diffusive timescale as a rate-limiting step, and hence the bound PSD only exhibits a single slope that is similar to the purely adsorbing case. At high frequencies and short timescales, the number of bound particles remains roughly constant, but the total number of particles and of free particles change due to diffusive entrance/exits effects. This results in a $f^{-3/2}$ scaling for both the free and total PSDs at high frequencies; the situation is similar to the purely diffusive case but with fewer diffusing particles, since some are bound. At low frequencies, the free and total PSDs show corners at the adsorption/desorption timescale, since at longer times particles can equilibrate across the whole system, diffusing, adsorbing, and desorbing.

In the context of Ref.~[\onlinecite{zevenbergen2009electrochemical}], their observation of $f^{-3/2}$ scaling at intermediate frequencies and $f^{-2}$ at high frequencies does not appear in this case in any of the measured fluctuating quantities. As such, we can conclude that the diffusive timescales must be slower than the adsorption/desorption timescales for their system. This highlights that the ordering of slopes in the PSD can be indicative of the relative timescales in the system, a relatively simple criterion to observe.

\subsubsection{Further Cases}

As highlighted in Table~\ref{tab:param_vals}, other experimental systems may exhibit further different timescale separations, where all timescales are comparable (Case 3) or where adsorption is much faster than all other processes (Case 4). In these scenarios, the spectra obtained are again non-linear combinations of the pure diffusion and pure adsorption cases, and can either be quite similar to either of the two pure cases (Case 3) or exhibit entirely peculiar shapes (Case 4). More detailed investigations are reported in Appendix~\ref{app:extraresults}. Overall, the two corners are not observed in these further cases.

\subsection{Key Takeaways}\label{sec:key_takeaways}

To summarize the different possibilities for the PSD shapes and scalings, we build a phase diagram -- see Fig.~\ref{fig:phase_diagram} -- comparing the diffusive and adsorption/desorption timescales.

The first key takeaway concerns PSD shapes. In most cases, the PSD appears to have an approximately Lorentzian, smooth, shape with a single well-defined or somewhat smeared out scaling regime that is more difficult to interpret. Yet, there exist a few special cases in which the PSDs can exhibit two corners, related to two distinct slopes. These special ``two corner'' cases are only observed in \textit{some but not all} of the spectra in the cases where there is a clear separation of timescales between the diffusive processes and adsorption/desorption processes (Cases 1 and 2). The PSD of free particles always exhibits the two corners in these cases, which makes sense since it is the only quantity for which both adsorption and diffusion affect number fluctuations directly. In contrast, the PSD of the total number of particles only exhibits two corners when diffusion is fast compared to binding, a condition by which it is indirectly able to sense the binding process. Reciprocally, the PSD of the number of bound particles only exhibits two corners when diffusion is slow compared to binding, a condition by which is is indirectly able to sense diffusion. 

The second key takeaway concerns the characteristic scalings that one can observe in the PSDs. The total and free PSDs always tend to a $f^{-3/2}$ scaling at high frequencies because they sense in particular particles that are diffusing and hence diffusive boundary crossings at short timescales. In contrast, the bound PSD always tends to a $f^{-2}$ scaling at high frequencies because it only senses adsorption and desorption processes to first order. In the cases (1 and 2) where there is a clear separation of timescales between diffusive and adsorption/desorption processes, two distinct scalings emerge (on the relevant curves) separated by a corner. These scalings appear necessarily in the order corresponding to the timescale separation: \textit{e.g.} if diffusion is the slower process, the $f^{-3/2}$ scaling will appear at intermediate frequencies, and $f^{-2}$ at high frequencies, and reciprocally if diffusion is the faster process. Naturally, because the free PSD has to decay to $f^{-3/2}$ at the highest frequencies but can exhibit a $f^{-2}$ scaling at high frequencies in the case of slow diffusion, the free PSD may exhibit two slopes at high frequencies separated by a slow transition, not a corner, which could mask the true scalings experimentally and result in effective fits of the slope between $-1.5$ to $-2.0$.

Finally, it is clear that there is significant non-additivity in these spectra and that, in general, there is not a straightforward route to construct the total PSD from the underlying spectra linked to pure diffusion and pure adsorption processes or from the bound and free PSDs. The physical reason for this is that processes are immediately coupled -- as soon as a particle desorbs it can diffuse -- and thus noise properties are convoluted.

\begin{figure}
    \centering
    \includegraphics[width=1\linewidth]{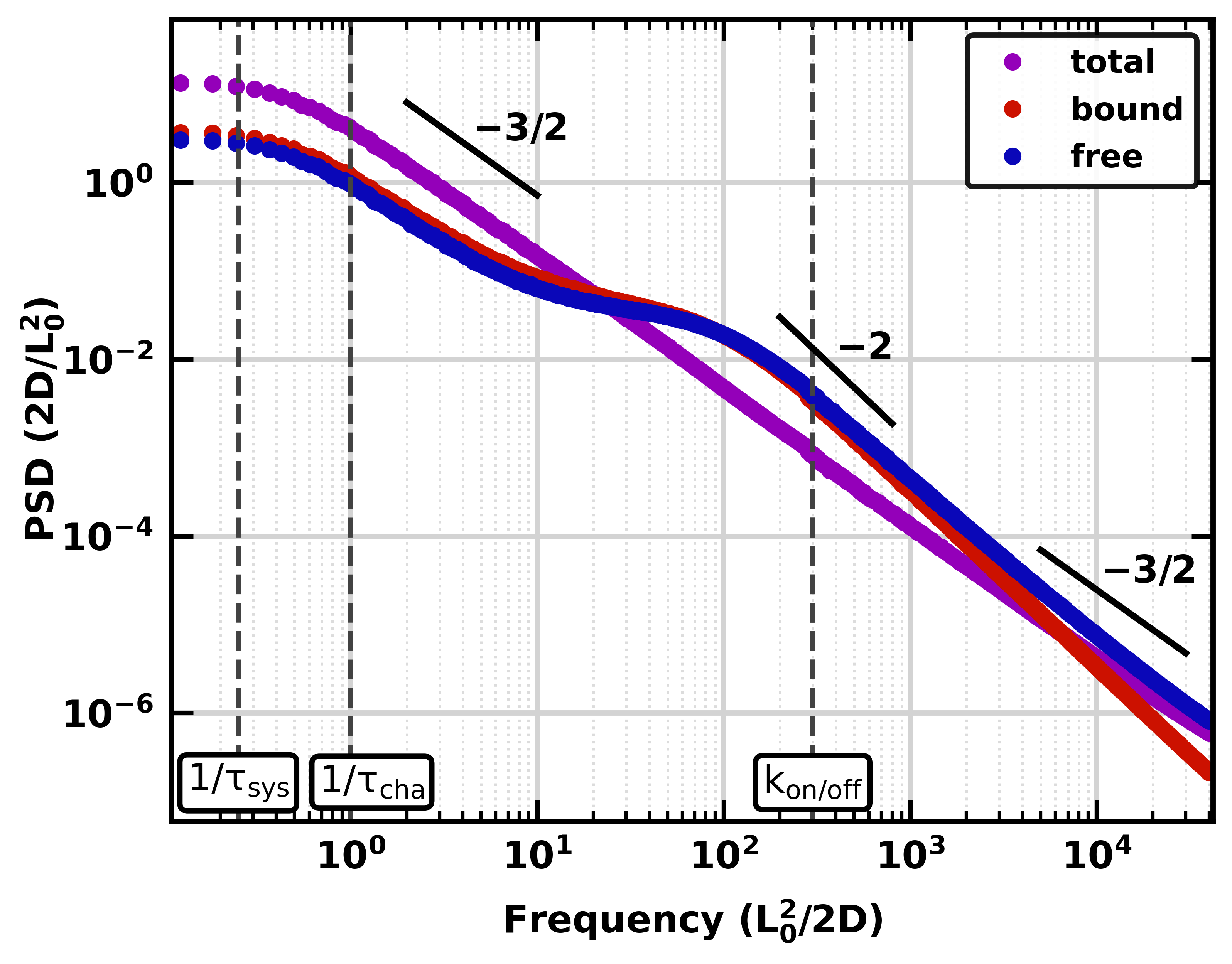}
    \caption{Simulated power spectral density of number fluctuations in the channel for a 3D system, for fluctuations in the total number (purple), the bound number (red), and the freely diffusing number (blue). We use $L=20\ell_0$ and $\kon=\koff=375/\tau_0$.}
    \label{fig:3dcase}
\end{figure}

More generally, we expect the phenomenology uncovered, in particular the presence of two corners or not according to the relative separation of timescales, should still hold in 3D or more complex geometries. To support this claim, we extend our simulations to a simple 3D system. As in Sec.~\ref{sec:pure_diffusion}, we run Brownian dynamics simulations in a 3D cube with an adsorbing region at the center. Results are shown in Fig.~\ref{fig:3dcase}. In this example, we choose the parameters to align with Case 1 of slow diffusion and fast adsorption/desorption. As in the 1D system, two slopes and two corner frequencies are observable in the bound and free PSDs due to the separation of timescales. Moreover, the slope below the corner corresponding to the adsorption/desorption timescales scales as $f^{-2}$, while at low frequencies below the diffusive timescales it is closer to $f^{-3/2}$. At high frequencies, the free PSD can be seen to tend to a $f^{-3/2}$ scaling. This is in agreement with our takeaways from the theoretical 1D case and demonstrates the robustness of our results to changes in geometry and dimension.

\section{Discussion and Conclusions}\label{sec:conc}

In this paper, we have derived expressions for the power spectrum of particle number fluctuations in a minimal confined system with adsorption and diffusion. By comparing our analytical results to simulated data, we have demonstrated that our expressions are valid over a wide range of parameters. We have investigated fluctuations in the number of bound, free, and total number of particles, finding remarkably different behavior across these quantities for a given parameter set. This highlights one of the main results of our investigation, namely, that the pure diffusion and pure adsorption components do not add up linearly in the PSD.

Spectra show a number of characteristic scalings that reflect key physical mechanisms and properties of the system, as outlined in Table~\ref{tab:scalings}. By considering the 1D case, we have also eliminated the effect of geometry on scaling \cite{gravelle2019adsorption,robin2023disentangling} and significantly simplified interpretation of the results. Interestingly, our analytic results indicate that there are only a small number of robust scalings which emerge directly from diffusive and adsorption mechanisms, with slopes of $f^{-3/2}$ and $f^{-2}$ respectively. Observations of other scalings are the result of intermediate transitions between regimes. The scaling of the power spectra thus also provides valuable insight into the dominant mechanism generating number fluctuations across different frequency ranges.

\begin{table}[h!]
    \centering
    \renewcommand{\arraystretch}{1.3}
    \setlength{\tabcolsep}{5pt}
    \begin{tabular}{ccp{0.5\linewidth}}
    \hline
        Scaling & Range & Mechanism\\
        \hline
        plateau & low $f$ & finite system size\\
        $1/f^{1/2}$ & intermediate $f$ & 1D re-entrances effects \\
        $1/f^{3/2}$ & high $f$ & diffusive crossings between regions \\
        $1/f^{2}$ & high $f$ & independent events like adsorption and desorption \\
        \hline
    \end{tabular}
    \caption{Key scalings in the system with their corresponding timescales or range of frequencies and physical mechanism.}
    \label{tab:scalings}
\end{table}

While our model is clearly simplistic, the diversity of results highlights how subtle and convoluted a PSD signal can be, and justifies the use of basic theoretical models to de-convolve PSD traces. Having at hand knowledge of when two corners and two slopes may be present, we can move on to investigate more complex systems. Such extensions could include more advanced geometries, which could be investigated for instance by adding another 1D line corresponding to the reservoirs \cite{marbach2021intrinsic}. Separately, we can carry out similar characterization of the 1D system after incorporation of other physical processes: for example, particle-particle interactions or a driving force in the channel. This would more accurately represent particles flowing through the channel under an applied electric field or pressure gradient, as is often the case in nanoporous or microfluidic experiments, and would open the door to probing non-equilibrium systems.

Real systems are clearly more complex than the situation considered here, but we can find scenarios in experiments that correspond to the different separations of timescales between diffusion and adsorption that we have studied. These are summarized in Fig.~\ref{fig:phase_diagram}, left-hand side, and clearly span the entire phase diagram, indicating that there is no general shape to be expected from a PSD even with such simple ingredients. The ordering of slopes can also be used in the first instance to determine relative timescales of diffusion and adsorption: a fairly simple criterion to check. An interesting question for all of these cases is which of our modeled PSDs -- total, free, or bound -- best reflects the physical quantity measured experimentally. For example, nanopore systems are often investigated via ionic current fluctuations. Here, the current is a measure of the ions in solution traveling \textit{through} the pore; therefore it could be reasonable to use the total PSD as an analogue for fluctuations in the current, assuming a degree of correlation between the pore conductance and total particles blocking the channel (adapter, sugar, etc) \cite{secchi2016scaling}. In nano-electrical devices, by design, the adsorbed particles may have the most significant effect on a macroscopic fluctuating signal, in which case the bound PSD is most relevant.

Our modeling does suggest a degree of caution in interpreting fluctuating behavior through only the knowledge of the PSD in experiments. In some cases, the difference between spectra are very small, \textit{e.g.} between a pure diffusion spectrum and the total or free PSDs when all timescales are comparable, or between a pure adsorption spectrum and the free PSD when adsorption is much faster than desorption. Differences between separate scenarios are clear when we know the correct parameters of the system to fit and compare to the theoretical results. In real experimental data, this is likely to be much harder to do, especially when two or more timescales are comparable, and the situation may become an inverse problem where several parameter combinations produce similar power spectra. Nonetheless, we have identified that the two corners can only appear under a specific separation of timescales suggesting that \textit{when} the two corners are seen, this means that one can in principle identify which regime of parameter space one is most likely in. This also suggests possible -- although likely non trivial -- experimental designs where \textit{e.g.} tuning the diffusive timescale may bring the system in one or another regime.

While our focus has predominantly been on investigating the PSD, it is clear that it is not the only way, and sometimes not even the relevant way, to investigate a fluctuating signal. In particular, because the PSD is generally obtained by averaging the signal over rolling time windows, it ``erases'' any memory effects in the signal. Other calculations on the fluctuating signal have revealed the non-Markovian nature of processes \cite{fulinski1998non} but are only rarely used. Given the current enthusiasm for memristive nanofluidic devices, which are based on long-time memory effects \cite{emmerich2024nanofluidic, robin2023long, kamsma2024brain}, it is likely that more diverse investigations of noise signals will soon be necessary.

\begin{acknowledgments}

The authors acknowledge fruitful discussions with Tristan Cerdin, Simon Gravelle and Pierre Levitz. The conception of this work was made possible thanks to numerous discussions with Aleksandar Donev. A. D. Y. is grateful to Max Howe for a very constructive discussion. S. M. received funding from the European Union’s Horizon 2020 research and innovation program under the Marie Skłodowska-Curie Grant Agreement No. 839225, MolecularControl. A. L. T. acknowledges funding from a Royal Society University Research Fellowship (URF/R1/211033) and from from EPSRC (EP/X02492X/1).

\end{acknowledgments}

\section*{Author Declarations}

\subsection*{Conflict of Interest}

The authors have no conflicts to disclose.

\subsection*{Author Contributions}

\textbf{Anna Drummond Young}: Conceptualization (equal); Investigation (lead); Methodology (equal); Software (lead); Visualization (equal); Writing - original draft (equal); Writing - review \& editing (equal). \textbf{Alice L. Thorneywork}: Conceptualization (equal); Funding acquisition (lead); Methodology (equal); Supervision (equal); Visualization (equal); Writing - original draft (equal); Writing - review \& editing (equal). \textbf{Sophie Marbach}: Conceptualization (equal); Investigation (equal); Methodology (equal); Supervision (equal); Visualization (equal); Writing - original draft (equal); Writing - review \& editing (equal).

\section*{Data Availability Statement}

The data that support the findings of this study are available within the article. To make the results in Eq.~\eqref{eqn:psd} more accessible, we have developed a sandbox where users can input their system parameters of choice and solve to find the resulting power spectra. The sandbox is in \href{https://colab.research.google.com/drive/15RnJ5kIYDBiprIDhEtW6KxADlR00WArb?usp=sharing}{Google Colab}, and available on GitHub: \href{https://github.com/ady-100/Noise-in-Nanofluidics}{\texttt{https://github.com/ady-100/Noise-in-Nanofluidics}}.

\appendix

\section{Solving the rates problem}\label{app:rates}

We designate $p_C$, $p_L$, and $p_R$ as representing variable $p$ in the channel (C) and left (L) and right (R) reservoirs, respectively. We define the Laplace transform $\hat{f}(x,s)$ of a function $f(x,t)$ as
\begin{equation}
    \mathcal{L}\{f\}(s) = \hat{f}(x,s) = \int_0^{\infty} e^{-st} f(x,t) dt.
    \label{eqn:laplacetransform}
\end{equation}
In Laplace space, a time derivative is equivalent to multiplication by $s$, such that $\mathcal{L}\{\partial_t p\}(s) = s \hat{p}(s) - p(x, 0)$.
This notably leads to the incorporation of the initial conditions \big($p(x, 0)$ and $q(x, 0)$\big) via the time derivative. Accordingly, we transform Eq.~\eqref{eqn:setupequations} into Laplace space, splitting $p$ up into $p_C$, $p_L$, and $p_R$ and using the initial conditions Eq.~\eqref{eqn:initialconditions},
\begin{equation}
\begin{cases}
    & \hat{p_C}=-\frac{\kon}{s}\hat{p_C}+\frac{\koff}{s}\hat{q}+\frac{D}{s}\partial_{xx}\hat{p_C} + \frac{p_0}{s}\\
    & \hat{q}=\frac{\kon}{s}\hat{p_C}-\frac{\koff}{s}\hat{q} + \frac{q_0}{s}\\
    & \hat{p_L}=\frac{D}{s}\partial_{xx}\hat{p_L}\\
    & \hat{p_R}=\frac{D}{s}\partial_{xx}\hat{p_R}
    \label{eqn:laplaceequations}
\end{cases}
\end{equation}
with boundary and continuity conditions
\begin{equation}
\begin{cases}
    &\mathrm{at} \,\, x=-L/2 \,\,\,\, \partial_x \hat{p_L} = 0 \\
    &\mathrm{at} \,\, x=-L_0/2 \,\,\,\,\displaystyle\begin{cases}\hat{p_L} =  \hat{p_C} \\
    \partial_x \hat{p_L} = \partial_x \hat{p_C}
    \end{cases} \\
    &\mathrm{at} \,\, x=L_0/2 \,\,\,\,\displaystyle\begin{cases}\hat{p_R} = \hat{p_C} \\
    \partial_x \hat{p_R} = \partial_x \hat{p_C}
    \end{cases} \\
    &\mathrm{at} \,\, x=L/2 \,\,\,\,\partial_x \hat{p_R} = 0.
    \label{eqn:laplaceboundaryconditions}
\end{cases}
\end{equation}

We solve the homogeneous equations for $\hat{p_L}$ and $\hat{p_R}$ by introducing the notation
\begin{equation}
    m_0(s) = \sqrt{\frac{s}{D}}
    \label{eqn:m0}
\end{equation}
to obtain
\begin{equation}
\begin{aligned}
    \hat{p_L} &= A_Le^{m_0x}+B_Le^{-m_0x} \\
    \hat{p_R} &= A_Re^{m_0x}+B_Re^{-m_0x}
    \label{eqn:plpr}
\end{aligned}
\end{equation}
where $A_{L/R}$ and $B_{L/R}$ are constants to be determined from the boundary conditions.
To solve for $\hat{p_C}$ and $\hat{q}$, we write $\hat{q}$ in terms of $\hat{p_C}$ using Eq.~\eqref{eqn:laplaceequations}
\begin{equation}
    \hat{q} = \frac{\kon \hat{p_C} + q_0}{s+\koff}
    \label{eqn:qasfuncofp}
\end{equation}
which allows us to simplify the equation for $\hat{p_C}$ into
\begin{equation}
    \partial_{xx}\hat{p_C} - \frac{s(s+\koff+\kon)}{D(s+\koff)} \hat{p_C} = -\frac{(\koff(p_0+q_0) + sp_0)}{D(s+\koff)}.
    \label{eqn:pcexpression}
\end{equation}
Introducing the notation
\begin{equation}
    m_1(s) = \sqrt{\frac{s(s+\koff+\kon)}{D(s+\koff)}}
    \label{eqn:m1}
\end{equation}
we find the solution to Eq.~\eqref{eqn:pcexpression}
\begin{equation}
    \hat{p_C} = A_Ce^{m_1x}+B_Ce^{-m_1x}+\frac{\koff(p_0+q_0) + sp_0}{s(s+\koff+\kon)}
    \label{eqn:pcresult}
\end{equation}
where again $A_C$ and $B_C$ are constants to be determined from the boundary conditions. 

Combining this with Eq.~\eqref{eqn:qasfuncofp} we find:
\begin{equation}
\begin{aligned}
    \hat{q} = \frac{\kon}{s+\koff}(A_Ce^{m_1x}&+B_Ce^{-m_1x})\\ +&\frac{\kon(p_0+q_0) + sq_0}{s(s+\koff+\kon)}.
    \label{eqn:qresult}
\end{aligned}
\end{equation}
There is a degree of symmetry between the results, and between the integration constants. Namely, using the boundary conditions in Eq.~\eqref{eqn:laplaceboundaryconditions}, we have
\normalsize\begin{equation}
\begin{aligned}    
    &A_C = B_C =  -\frac{m_0 (\koff(p_0+q_0) + p_0 s)}{s(s+\koff+\kon)} \times ... \\
    &\frac{e^{L_0 m_1/2} (e^{L m_0} - e^{L_0 m_0})}{(m_0-m_1)(e^{L m_0} - e^{L_0(m_0+m_1)}) + (m_0+m_1)(e^{L m_0 + L_0 m_1} - e^{L_0 m_0})}.
    \label{eqn:acbc}
\end{aligned}
\end{equation}
\normalsize
We do not report the expressions of the other integration constants, since they will not be required in the following steps.

\section{Further correlation functions}\label{app:othercorrelations}

\subsubsection{Number of adsorbed particles}

Following the same method as Sec.~\ref{sec:numberflucs}, we calculate the correlation function for the number of bound particles in the channel
\begin{equation}
\begin{aligned}
    C_{\rm on}(t) &= \langle N_{\rm on}(t)N_{\rm on}(0) \rangle \\
    &= \mathcal{N} \psi_{\rm on\rightarrow on}(t)
    \label{eqn:boundparticlesprobs}
\end{aligned}
\end{equation}
where $\psi_{\rm on\rightarrow on}(t)$ represents the probability that a particle \textit{starting adsorbed} in the channel is \textit{still adsorbed} in the channel at time $t$. Importantly, this means that the initial conditions in Eq.~\eqref{eqn:initialconditions} now become $p_0=0$. We have
\begin{equation}
    \psi_{\rm on\rightarrow on}(t) =  \int_{-L_0/2}^{L_0/2} q(x,t) dx.
    \label{eqn:psion}
\end{equation}
We take, as for the total number, the time derivative and Laplace transform and using Eqs.~\eqref{eqn:pcresult} and \eqref{eqn:qresult}, we obtain:
\begin{equation}
\begin{aligned}
    \hat{C}_{\rm on}&(s) = 2\mathcal{N} \Bigg[ \frac{\kon A_C}{m_1(s+\koff)} \\ &\Big(e^{L_0 m_1/2} - e^{-L_0 m_1/2}\Big) - \frac{q_0 \koff L_0}{2s(s+\koff+\kon)} \Bigg].
    \label{eqn:correlationfuncon}
\end{aligned}
\end{equation}

\subsubsection{Number of unadsorbed particles}

Again we follow the same method as Sec.~\ref{sec:numberflucs}. We have:
\begin{equation}
\begin{aligned}
    C_{\rm off}(t) &= \langle N_{\rm off}(t)N_{\rm off}(0) \rangle \\
    &= \mathcal{N} \psi_{\rm off\rightarrow off}(t)
    \label{eqn:unboundparticlesprobs}
\end{aligned}
\end{equation}
where this time $\psi_{\rm off\rightarrow off}(t)$ represents the probability that a particle \textit{starting unadsorbed} in the channel is \textit{still unadsorbed} in the channel at time $t$. Importantly, this means that the initial conditions now include $q_0=0$. We have:
\begin{equation}
    \psi_{\rm off\rightarrow off}(t) =  \int_{-L_0/2}^{L_0/2} p(x,t) dx.
    \label{eqn:psioff}
\end{equation}
We take, as for the total number, the time derivative and Laplace transform and using Eqs.~\eqref{eqn:pcresult} and \eqref{eqn:qresult}, we obtain:
\begin{equation}
\begin{aligned}
    \hat{C}_{\rm off}&(s) = -2\mathcal{N} \Bigg[A_C \Big(e^{L_0 m_1 /2} - e^{-L_0 m_1 /2}\Big) \\
    &\Big(\frac{\kon}{m_1(s+\koff)} - \frac{m_1}{m_0^2}\Big) + \frac{p_0 \kon L_0}{2 s (s + \koff + \kon)} \Bigg].
    \label{eqn:correlationfuncoff}
\end{aligned}
\end{equation}

\subsubsection{Cross-correlations}

In general, we can also calculate correlation functions for the numbers of particles which switch from bound to free, $C_{\rm on\rightarrow off}(t)$, and vice versa, $C_{\rm off\rightarrow on}(t)$. These cross-correlations are non-trivial, which means that in most cases one cannot simply sum the contributions to the total PSD of free and bound number fluctuations, i.e. $S_{N}(f) \ne S_{\rm on}(f) + S_{\rm on}(f)$. The cross-correlations can be calculated using the same method as Sec.~\ref{sec:numberflucs}:
\begin{equation}
\begin{aligned}
    C_{\rm on\rightarrow off}(t) &= \langle N_{\rm off}(t)N_{\rm on}(0) \rangle \\
    &= \mathcal{N} \psi_{\rm on\rightarrow off}(t)
\end{aligned}
\end{equation}
where $\psi_{\rm on\rightarrow off}(t)$ represents the probability that a particle \textit{starting adsorbed} in the channel is \textit{unadsorbed} in the channel at time $t$. The initial condition is now that $p_0=0$. Again we take the time derivative and Laplace transform and using Eqs.~\eqref{eqn:pcresult} and \eqref{eqn:qresult}, we obtain:
\begin{equation}
\begin{aligned}
    \hat{C}_{\rm on\rightarrow off}&(s) = 2\mathcal{N} \Bigg[A_C \Big(e^{L_0 m_1 /2} - e^{-L_0 m_1 /2}\Big) \\
    &\Big(\frac{m_1}{m_0^2} - \frac{\kon}{m_1(s+\koff)} \Big) + \frac{q_0 \koff L_0}{2 s (s + \koff + \kon)} \Bigg].
\end{aligned}
\end{equation}
The other cross-correlations follows:
\begin{equation}
\begin{aligned}
    C_{\rm off\rightarrow on}(t) &= \langle N_{\rm on}(t)N_{\rm off}(0) \rangle \\
    &= \mathcal{N} \psi_{\rm off\rightarrow on}(t)
\end{aligned}
\end{equation}
now with $q_0=0$. As above:
\begin{equation}
\begin{aligned}
    \hat{C}_{\rm off\rightarrow on}&(s) = 2\mathcal{N} \Bigg[A_C \Big(e^{L_0 m_1 /2} - e^{-L_0 m_1 /2}\Big) \\
    &\Big(\frac{\kon}{m_1(s+\koff)} \Big) + \frac{p_0 \kon L_0}{2 s (s + \koff + \kon)} \Bigg].
\end{aligned}
\end{equation}

\section{Computational methods}\label{app:methods}

The system we simulate is that described in Fig.~\ref{fig:theorysetup}. Particles are initialized along the lines $P$ and $Q$ with an initial distribution calculated from the expected number of particles in each region according to the probability densities $p_0$ and $q_0$. This allows the steady state to be reached faster than random initialization. The system contains $\mathcal{N}$ particles and $N_{\rm T}$ steps are carried out in total. We use $\mathcal{N}=100$ and $N_{\rm T}=2 \times 10^5$. The time step used is $dt$, and is chosen to be sufficiently small compared to the timescales of diffusion, adsorption, and desorption that it captures their full behavior, \textit{i.e.}:
\begin{equation}
\begin{aligned}
    dt &\ll 1/\kon \\
    dt &\ll 1/\koff \\
    dt &\ll L_0^2/D.
    \label{eqn:dtconstraint}
\end{aligned}
\end{equation}
The probability for a given particle inside the channel to adsorb in a time step is $p_{\rm on}=\kon dt$ and the probability to desorb is $p_{\rm off}=\koff dt$.

Each step is carried out as follows. For each particle, we first check if it is already adsorbed. If it is, a random number is sampled from a uniform distribution $r_1\sim U(0, 1)$. If $r_1<p_{\rm off}$, the particle un-adsorbs, and if $r_1>p_{\rm off}$, it remains adsorbed. If the particle was not previously adsorbed, but is in the channel region, a second random number $r_2\sim U(0, 1)$ is generated. If $r_2<p_{\rm on}$, the particle adsorbs, and if $r_2>p_{\rm on}$, it remains unadsorbed. If it remains unadsorbed, or if it was initially in the reservoirs, the particle moves according to:
\begin{equation}
    X(t+dt)=X(t)+\sqrt{2Ddt} \, \mathcal{W}_t
    \label{eqn:stepdisplacement}
\end{equation}
where $\mathcal{W}_t$ is a sampled from a normal distribution at each time step, $\mathcal{W}_t\sim N(0, 1)$. If $X(t+dt)$ lies inside the system it is accepted; however, if it lies outside the allowed region it is reflected back into the allowed region according to Snell's law (specular reflection)~\cite{scala2007event}.

\begin{figure*}
    \centering
    \includegraphics[width=1\textwidth]{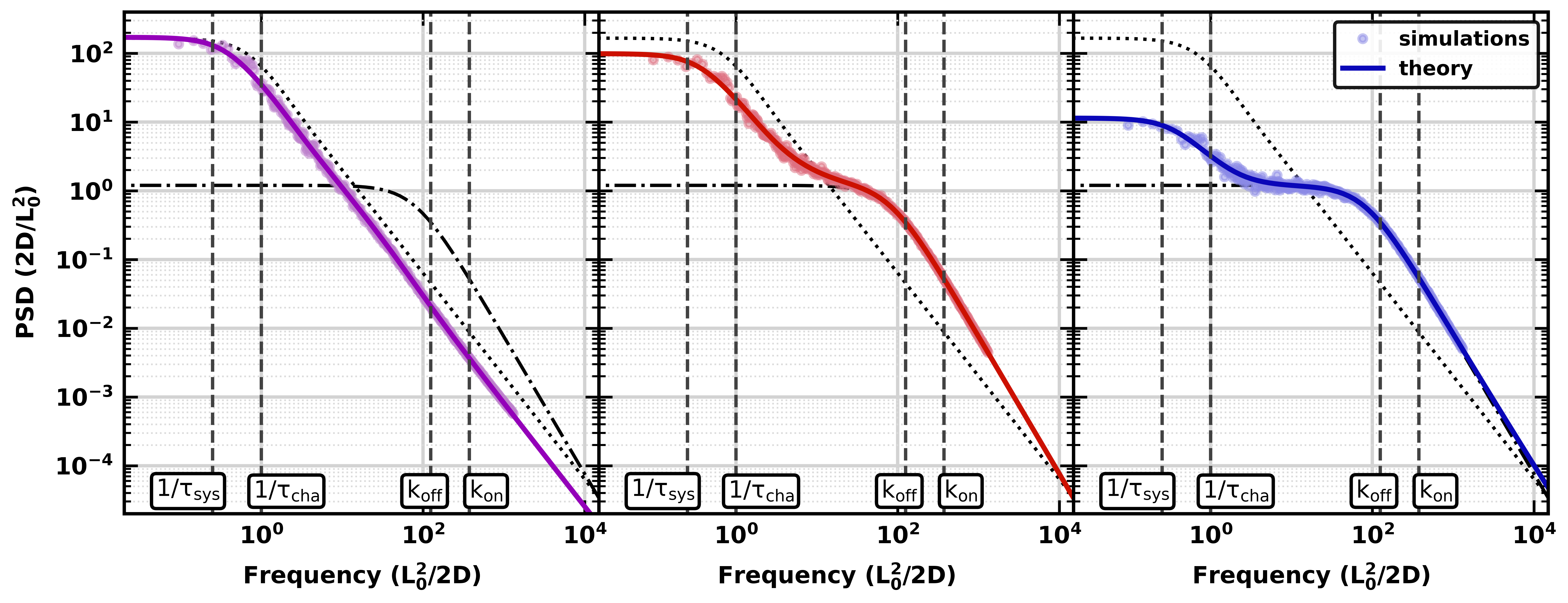}
    \caption{\textbf{Slow diffusion, fast adsorption/desorption -- Case 1 (continued): $\tau_{\rm cha} \gg \tau_{\rm ads} \sim \tau_{\rm des}$.} Left: total PSD. Middle: bound PSD. Right: free PSD. We use $L=2\ell_0$, $\kon=375/\tau_0$, and $\koff=125/\tau_0$. Note that the main difference compared to Fig.~\ref{fig:case_1a} is that relative ordering of the rates of adsorption and desorption are now reversed, here $\kon \gg \koff$. Labels are similar to Fig.~\ref{fig:case_1a}.}
    \label{fig:case_1b}
\end{figure*}

We measure the number of particles in the channel, or specific region of the channel, at every time step. To characterize these fluctuations, we calculate the PSD which can be defined as:
\begin{equation}
\begin{aligned}
    S(f) = |\mathcal{F}\{x(t)\}|^2
    \label{eqn:psdfromFT}
\end{aligned}
\end{equation}
where $x(t)$ is the time-varying signal and $\mathcal{F}$ represents a Fourier transform. Since we are measuring a discrete signal -- the particle number fluctuations -- we cannot apply Eq.~\eqref{eqn:psdfromFT} directly. We instead estimate the power spectrum using Welch's method \cite{welch2003use}, which takes the average PSD from multiple overlapping sections of the signal. This reduces statistical error on $S(f)$, an advantage over other methods of PSD estimation. To further reduce error, we run each simulation $10$ times and average the estimated PSDs. The spectrum is then smoothed by logarithmically binning along the frequency axis and averaging $S(f)$ in each bin. Then the PSD of the simulated number fluctuations can be compared directly to the theoretical results.

In the 3D simulations, the procedure is exactly as above, except each particle has three position coordinates and diffuses in a cubic box of side length $L$. The ``channel region'', or sensing region, is a cubic box of length $L_0$ at the center.

\section{Additional results in the full case}\label{app:extraresults}

Here we consider additional representative results in the full case with adsorption processes and diffusion.

\begin{figure*}
    \centering
    \includegraphics[width=1\textwidth]{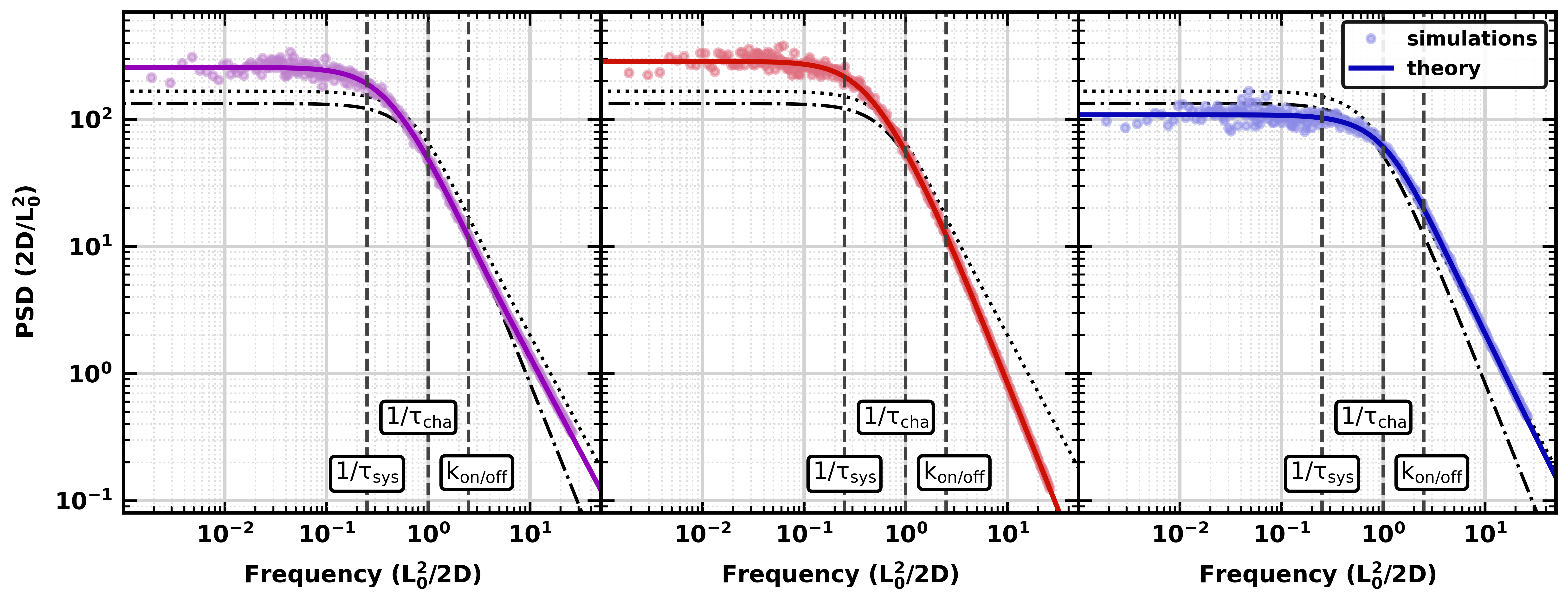}
    \caption{\textbf{All physical processes have comparable timescales -- Case 3: $\tau_{\rm cha} \sim \tau_{\rm ads} \sim \tau_{\rm des}$.} Left: total PSD. Middle: bound PSD. Right: free PSD. We use $L=2 \ell_0$ and $\kon=\koff=2.5/\tau_0$.}
    \label{fig:case_3}
\end{figure*}

\subsection{Diffusion Slower than Adsorption/Desorption}

First we consider the complementary case to that in Sec.~\ref{sec:resultsDSAD}. Changing the ordering of the adsorption and desorption timescales does not significantly affect the shape of the power spectra, but nonetheless there are some subtle changes. Results are in Fig.~\ref{fig:case_1b}.

Now, when the adsorption rate is faster than desorption and binding is favored, the total PSD lies further from the purely diffusive line (dotted) at high frequencies. In this region the free PSD lies closer to the pure adsorption line (dash-dot). These changes can be reconciled with the fact that increasing the adsorption rate makes binding more favorable, so particles adsorb more quickly on entering the channel, on average. This suppresses the high frequency $f^{-3/2}$ scaling due to the diffusive motion of particles across boundaries in the total PSD, and also increases the importance of the adsorption process to the free PSD. Nonetheless, at very high frequencies both toal and free PSDs eventually recover their $f^{-3/2}$ scaling.

\subsection{Comparable Timescales}

Spectra for the case where diffusion, adsorption, and desorption all have similar timescales are shown in Fig.~\ref{fig:case_3}. Here, the two corner frequencies and intermediate scaling we could observe clearly in the previous two cases collapse, with all three PSDs showing only one plateau and corner frequency. This illustrates how a complex physical situation with overlapping timescales could be difficult to interpret experimentally. While there are four timescales and two distinct processes (adsorption and diffusion) in the system, they are masked by the simple shape of the PSD.

However, our model and simulations provide a possible means of distinguishing this case from the similar-looking cases of pure diffusion and pure adsorption. The total PSD is close to the pure diffusive case but at intermediate frequencies transiently follows the adsorption case with a scaling of $f^{-2}$. The bound PSD, while similar to the case of pure adsorption, has a higher plateau. The free PSD resembles the case of pure diffusion, but its slope is steeper than $f^{-3/2}$ at intermediate frequencies as it is still affected by adsorption (until it recovers the $f^{-3/2}$ scaling at high frequencies.

Where could we observe this kind of behavior? One example is in the adsorption of polymers inside glass nanopores \cite{knowles2021current}. We could also see an overlap in diffusion and adsorption in DNA hybridization microarray assays \cite{gadgil2004diffusion}. Although diffusion of small molecules is generally ``fast'', the diffusive timescale becomes long when the channel length involved increases to hundreds of ${\rm nm}$ or even ${\rm \mu m}$.

\begin{figure*}
    \centering
    \includegraphics[width=1\textwidth]{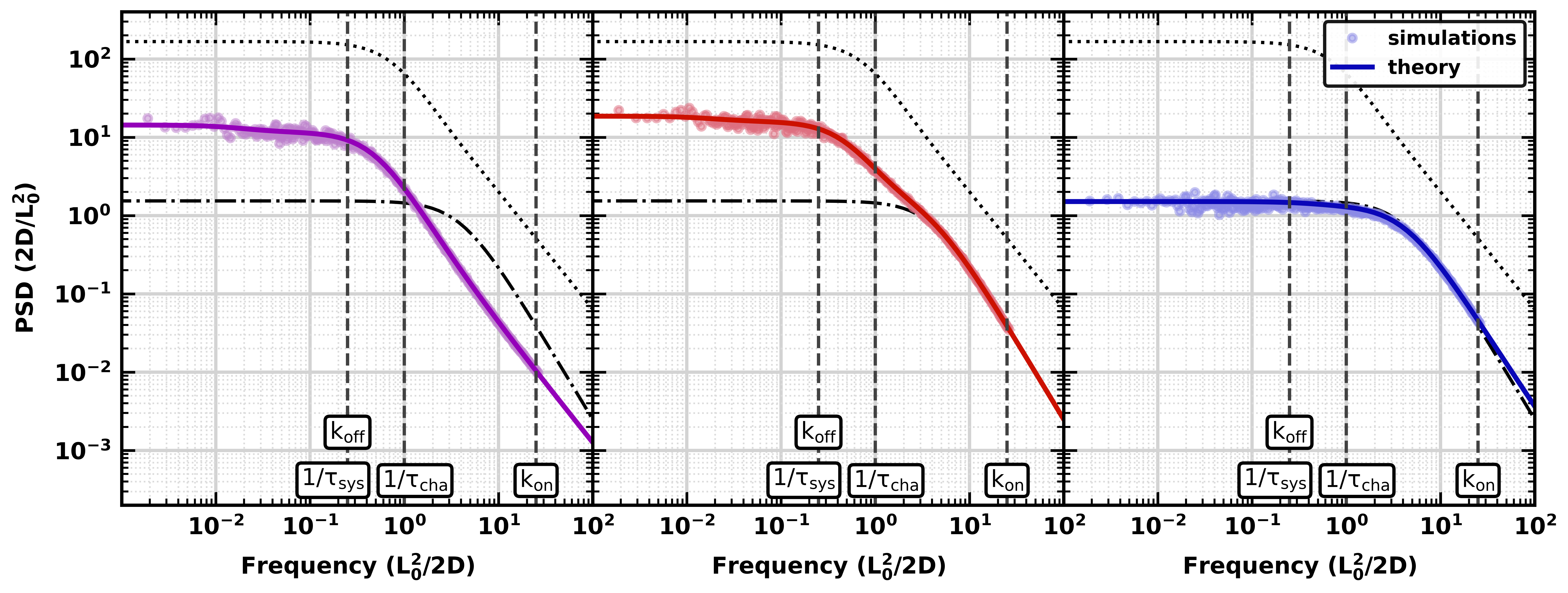}
    \caption{\textbf{Adsorption much faster than desorption -- Case 4: $\tau_{\rm des} \gg \tau_{\rm sys} \sim \tau_{\rm cha} \gg \tau_{\rm ads}$.} Left: total PSD. Middle: bound PSD. Right: free PSD. We use $L=2\ell_0$, $\kon=25/\tau_0$, and $\koff=0.25/\tau_0$.}
    \label{fig:case_4}
\end{figure*}

\subsection{Separation of Adsorption and Desorption Timescales}

Lastly, we consider the case of the separation of adsorption and desorption timescales. When desorption is much faster than adsorption, we tend to the pure diffusive case, so we do not consider that here. However, when adsorption is the faster process, we have a more interesting case where particles enter the channel and immediately adsorb. This means particles effectively transition between the reservoirs and the bound state. Although this may seem like an extreme case, this situation can occur physically in any system where adsorption is heavily favored. This is the case in the catalysis of benzene alkylation in zeolites \cite{zhao2016flow}, as well as in microporous biomolecule sensors \cite{hansen2009analysis}.

Representative results are shown in Fig.~\ref{fig:case_4}. The PSDs now have more complex shapes. The total PSD shows several scaling regimes, with a corner at the diffusive timescale and a kink at the adsorption timescale. Above the adsorption timescale, it scales as $f^{-3/2}$. The bound PSD also shows at least two separate scalings and corner frequencies. At high frequencies, it resembles the pure adsorption case with a $f^{-2}$ scaling. The free PSD is very close to the case of pure adsorption (telegraph noise) but its slope at high frequencies eventually returns to $f^{-3/2}$. Interestingly, the diffusive timescale along the channel is the only timescale not corresponding to a change in slope for these PSDs - this may be because very few (if any) particles are able to diffuse along the channel without adsorption.

\nocite{*}
\section*{References}


\begin{thebibliography}{72}%
\makeatletter
\providecommand \@ifxundefined [1]{%
 \@ifx{#1\undefined}
}%
\providecommand \@ifnum [1]{%
 \ifnum #1\expandafter \@firstoftwo
 \else \expandafter \@secondoftwo
 \fi
}%
\providecommand \@ifx [1]{%
 \ifx #1\expandafter \@firstoftwo
 \else \expandafter \@secondoftwo
 \fi
}%
\providecommand \natexlab [1]{#1}%
\providecommand \enquote  [1]{``#1''}%
\providecommand \bibnamefont  [1]{#1}%
\providecommand \bibfnamefont [1]{#1}%
\providecommand \citenamefont [1]{#1}%
\providecommand \href@noop [0]{\@secondoftwo}%
\providecommand \href [0]{\begingroup \@sanitize@url \@href}%
\providecommand \@href[1]{\@@startlink{#1}\@@href}%
\providecommand \@@href[1]{\endgroup#1\@@endlink}%
\providecommand \@sanitize@url [0]{\catcode `\\12\catcode `\$12\catcode `\&12\catcode `\#12\catcode `\^12\catcode `\_12\catcode `\%12\relax}%
\providecommand \@@startlink[1]{}%
\providecommand \@@endlink[0]{}%
\providecommand \url  [0]{\begingroup\@sanitize@url \@url }%
\providecommand \@url [1]{\endgroup\@href {#1}{\urlprefix }}%
\providecommand \urlprefix  [0]{URL }%
\providecommand \Eprint [0]{\href }%
\providecommand \doibase [0]{http://dx.doi.org/}%
\providecommand \selectlanguage [0]{\@gobble}%
\providecommand \bibinfo  [0]{\@secondoftwo}%
\providecommand \bibfield  [0]{\@secondoftwo}%
\providecommand \translation [1]{[#1]}%
\providecommand \BibitemOpen [0]{}%
\providecommand \bibitemStop [0]{}%
\providecommand \bibitemNoStop [0]{.\EOS\space}%
\providecommand \EOS [0]{\spacefactor3000\relax}%
\providecommand \BibitemShut  [1]{\csname bibitem#1\endcsname}%
\let\auto@bib@innerbib\@empty
\bibitem [{\citenamefont {Neher}\ and\ \citenamefont {Sakmann}(1976)}]{neher1976single}%
  \BibitemOpen
  \bibfield  {author} {\bibinfo {author} {\bibfnamefont {E.}~\bibnamefont {Neher}}\ and\ \bibinfo {author} {\bibfnamefont {B.}~\bibnamefont {Sakmann}},\ }\bibfield  {title} {\enquote {\bibinfo {title} {Single-channel currents recorded from membrane of denervated frog muscle fibres},}\ }\href@noop {} {\bibfield  {journal} {\bibinfo  {journal} {Nature}\ }\textbf {\bibinfo {volume} {260}},\ \bibinfo {pages} {799--802} (\bibinfo {year} {1976})}\BibitemShut {NoStop}%
\bibitem [{\citenamefont {Cooper}, \citenamefont {Jakobsson},\ and\ \citenamefont {Wolynes}(1985)}]{cooper1985theory}%
  \BibitemOpen
  \bibfield  {author} {\bibinfo {author} {\bibfnamefont {K.}~\bibnamefont {Cooper}}, \bibinfo {author} {\bibfnamefont {E.}~\bibnamefont {Jakobsson}}, \ and\ \bibinfo {author} {\bibfnamefont {P.}~\bibnamefont {Wolynes}},\ }\bibfield  {title} {\enquote {\bibinfo {title} {The theory of ion transport through membrane channels},}\ }\href@noop {} {\bibfield  {journal} {\bibinfo  {journal} {Progress in biophysics and molecular biology}\ }\textbf {\bibinfo {volume} {46}},\ \bibinfo {pages} {51--96} (\bibinfo {year} {1985})}\BibitemShut {NoStop}%
\bibitem [{\citenamefont {Nikaido}(1992)}]{nikaido1992porins}%
  \BibitemOpen
  \bibfield  {author} {\bibinfo {author} {\bibfnamefont {H.}~\bibnamefont {Nikaido}},\ }\bibfield  {title} {\enquote {\bibinfo {title} {Porins and specific channels of bacterial outer membranes},}\ }\href@noop {} {\bibfield  {journal} {\bibinfo  {journal} {Molecular microbiology}\ }\textbf {\bibinfo {volume} {6}},\ \bibinfo {pages} {435--442} (\bibinfo {year} {1992})}\BibitemShut {NoStop}%
\bibitem [{\citenamefont {Kasianowicz}\ \emph {et~al.}(1996)\citenamefont {Kasianowicz}, \citenamefont {Brandin}, \citenamefont {Branton},\ and\ \citenamefont {Deamer}}]{kasianowicz1996characterization}%
  \BibitemOpen
  \bibfield  {author} {\bibinfo {author} {\bibfnamefont {J.~J.}\ \bibnamefont {Kasianowicz}}, \bibinfo {author} {\bibfnamefont {E.}~\bibnamefont {Brandin}}, \bibinfo {author} {\bibfnamefont {D.}~\bibnamefont {Branton}}, \ and\ \bibinfo {author} {\bibfnamefont {D.~W.}\ \bibnamefont {Deamer}},\ }\bibfield  {title} {\enquote {\bibinfo {title} {Characterization of individual polynucleotide molecules using a membrane channel},}\ }\href@noop {} {\bibfield  {journal} {\bibinfo  {journal} {Proceedings of the National Academy of Sciences}\ }\textbf {\bibinfo {volume} {93}},\ \bibinfo {pages} {13770--13773} (\bibinfo {year} {1996})}\BibitemShut {NoStop}%
\bibitem [{\citenamefont {Clarke}\ \emph {et~al.}(2009)\citenamefont {Clarke}, \citenamefont {Wu}, \citenamefont {Jayasinghe}, \citenamefont {Patel}, \citenamefont {Reid},\ and\ \citenamefont {Bayley}}]{clarke2009continuous}%
  \BibitemOpen
  \bibfield  {author} {\bibinfo {author} {\bibfnamefont {J.}~\bibnamefont {Clarke}}, \bibinfo {author} {\bibfnamefont {H.-C.}\ \bibnamefont {Wu}}, \bibinfo {author} {\bibfnamefont {L.}~\bibnamefont {Jayasinghe}}, \bibinfo {author} {\bibfnamefont {A.}~\bibnamefont {Patel}}, \bibinfo {author} {\bibfnamefont {S.}~\bibnamefont {Reid}}, \ and\ \bibinfo {author} {\bibfnamefont {H.}~\bibnamefont {Bayley}},\ }\bibfield  {title} {\enquote {\bibinfo {title} {Continuous base identification for single-molecule nanopore dna sequencing},}\ }\href@noop {} {\bibfield  {journal} {\bibinfo  {journal} {Nature nanotechnology}\ }\textbf {\bibinfo {volume} {4}},\ \bibinfo {pages} {265--270} (\bibinfo {year} {2009})}\BibitemShut {NoStop}%
\bibitem [{\citenamefont {Bell}\ and\ \citenamefont {Keyser}(2016)}]{bell2016digitally}%
  \BibitemOpen
  \bibfield  {author} {\bibinfo {author} {\bibfnamefont {N.~A.}\ \bibnamefont {Bell}}\ and\ \bibinfo {author} {\bibfnamefont {U.~F.}\ \bibnamefont {Keyser}},\ }\bibfield  {title} {\enquote {\bibinfo {title} {Digitally encoded dna nanostructures for multiplexed, single-molecule protein sensing with nanopores},}\ }\href@noop {} {\bibfield  {journal} {\bibinfo  {journal} {Nature nanotechnology}\ }\textbf {\bibinfo {volume} {11}},\ \bibinfo {pages} {645--651} (\bibinfo {year} {2016})}\BibitemShut {NoStop}%
\bibitem [{\citenamefont {Deamer}, \citenamefont {Akeson},\ and\ \citenamefont {Branton}(2016)}]{deamer2016three}%
  \BibitemOpen
  \bibfield  {author} {\bibinfo {author} {\bibfnamefont {D.}~\bibnamefont {Deamer}}, \bibinfo {author} {\bibfnamefont {M.}~\bibnamefont {Akeson}}, \ and\ \bibinfo {author} {\bibfnamefont {D.}~\bibnamefont {Branton}},\ }\bibfield  {title} {\enquote {\bibinfo {title} {Three decades of nanopore sequencing},}\ }\href@noop {} {\bibfield  {journal} {\bibinfo  {journal} {Nature biotechnology}\ }\textbf {\bibinfo {volume} {34}},\ \bibinfo {pages} {518--524} (\bibinfo {year} {2016})}\BibitemShut {NoStop}%
\bibitem [{\citenamefont {Werber}, \citenamefont {Osuji},\ and\ \citenamefont {Elimelech}(2016)}]{werber2016materials}%
  \BibitemOpen
  \bibfield  {author} {\bibinfo {author} {\bibfnamefont {J.~R.}\ \bibnamefont {Werber}}, \bibinfo {author} {\bibfnamefont {C.~O.}\ \bibnamefont {Osuji}}, \ and\ \bibinfo {author} {\bibfnamefont {M.}~\bibnamefont {Elimelech}},\ }\bibfield  {title} {\enquote {\bibinfo {title} {Materials for next-generation desalination and water purification membranes},}\ }\href@noop {} {\bibfield  {journal} {\bibinfo  {journal} {Nature Reviews Materials}\ }\textbf {\bibinfo {volume} {1}},\ \bibinfo {pages} {1--15} (\bibinfo {year} {2016})}\BibitemShut {NoStop}%
\bibitem [{\citenamefont {Siria}, \citenamefont {Bocquet},\ and\ \citenamefont {Bocquet}(2017)}]{siria2017new}%
  \BibitemOpen
  \bibfield  {author} {\bibinfo {author} {\bibfnamefont {A.}~\bibnamefont {Siria}}, \bibinfo {author} {\bibfnamefont {M.-L.}\ \bibnamefont {Bocquet}}, \ and\ \bibinfo {author} {\bibfnamefont {L.}~\bibnamefont {Bocquet}},\ }\bibfield  {title} {\enquote {\bibinfo {title} {New avenues for the large-scale harvesting of blue energy},}\ }\href@noop {} {\bibfield  {journal} {\bibinfo  {journal} {Nature Reviews Chemistry}\ }\textbf {\bibinfo {volume} {1}},\ \bibinfo {pages} {0091} (\bibinfo {year} {2017})}\BibitemShut {NoStop}%
\bibitem [{\citenamefont {Ying}\ \emph {et~al.}(2022)\citenamefont {Ying}, \citenamefont {Hu}, \citenamefont {Zhang}, \citenamefont {Qing}, \citenamefont {Fragasso}, \citenamefont {Maglia}, \citenamefont {Meller}, \citenamefont {Bayley}, \citenamefont {Dekker},\ and\ \citenamefont {Long}}]{ying2022nanopore}%
  \BibitemOpen
  \bibfield  {author} {\bibinfo {author} {\bibfnamefont {Y.-L.}\ \bibnamefont {Ying}}, \bibinfo {author} {\bibfnamefont {Z.-L.}\ \bibnamefont {Hu}}, \bibinfo {author} {\bibfnamefont {S.}~\bibnamefont {Zhang}}, \bibinfo {author} {\bibfnamefont {Y.}~\bibnamefont {Qing}}, \bibinfo {author} {\bibfnamefont {A.}~\bibnamefont {Fragasso}}, \bibinfo {author} {\bibfnamefont {G.}~\bibnamefont {Maglia}}, \bibinfo {author} {\bibfnamefont {A.}~\bibnamefont {Meller}}, \bibinfo {author} {\bibfnamefont {H.}~\bibnamefont {Bayley}}, \bibinfo {author} {\bibfnamefont {C.}~\bibnamefont {Dekker}}, \ and\ \bibinfo {author} {\bibfnamefont {Y.-T.}\ \bibnamefont {Long}},\ }\bibfield  {title} {\enquote {\bibinfo {title} {Nanopore-based technologies beyond dna sequencing},}\ }\href@noop {} {\bibfield  {journal} {\bibinfo  {journal} {Nature nanotechnology}\ }\textbf {\bibinfo {volume} {17}},\ \bibinfo {pages} {1136--1146} (\bibinfo {year} {2022})}\BibitemShut {NoStop}%
\bibitem [{\citenamefont {Dekker}(2007)}]{dekker2007solid}%
  \BibitemOpen
  \bibfield  {author} {\bibinfo {author} {\bibfnamefont {C.}~\bibnamefont {Dekker}},\ }\bibfield  {title} {\enquote {\bibinfo {title} {Solid-state nanopores},}\ }\href@noop {} {\bibfield  {journal} {\bibinfo  {journal} {Nature nanotechnology}\ }\textbf {\bibinfo {volume} {2}},\ \bibinfo {pages} {209--215} (\bibinfo {year} {2007})}\BibitemShut {NoStop}%
\bibitem [{\citenamefont {Goto}\ \emph {et~al.}(2020)\citenamefont {Goto}, \citenamefont {Akahori}, \citenamefont {Yanagi},\ and\ \citenamefont {Takeda}}]{goto2020solid}%
  \BibitemOpen
  \bibfield  {author} {\bibinfo {author} {\bibfnamefont {Y.}~\bibnamefont {Goto}}, \bibinfo {author} {\bibfnamefont {R.}~\bibnamefont {Akahori}}, \bibinfo {author} {\bibfnamefont {I.}~\bibnamefont {Yanagi}}, \ and\ \bibinfo {author} {\bibfnamefont {K.-i.}\ \bibnamefont {Takeda}},\ }\bibfield  {title} {\enquote {\bibinfo {title} {Solid-state nanopores towards single-molecule dna sequencing},}\ }\href@noop {} {\bibfield  {journal} {\bibinfo  {journal} {Journal of human genetics}\ }\textbf {\bibinfo {volume} {65}},\ \bibinfo {pages} {69--77} (\bibinfo {year} {2020})}\BibitemShut {NoStop}%
\bibitem [{\citenamefont {Siwy}, \citenamefont {Bruening},\ and\ \citenamefont {Howorka}(2023)}]{siwy2023nanopores}%
  \BibitemOpen
  \bibfield  {author} {\bibinfo {author} {\bibfnamefont {Z.~S.}\ \bibnamefont {Siwy}}, \bibinfo {author} {\bibfnamefont {M.~L.}\ \bibnamefont {Bruening}}, \ and\ \bibinfo {author} {\bibfnamefont {S.}~\bibnamefont {Howorka}},\ }\bibfield  {title} {\enquote {\bibinfo {title} {Nanopores: synergy from dna sequencing to industrial filtration--small holes with big impact},}\ }\href@noop {} {\bibfield  {journal} {\bibinfo  {journal} {Chemical Society Reviews}\ }\textbf {\bibinfo {volume} {52}},\ \bibinfo {pages} {1983--1994} (\bibinfo {year} {2023})}\BibitemShut {NoStop}%
\bibitem [{\citenamefont {Chazot-Franguiadakis}\ \emph {et~al.}(2022)\citenamefont {Chazot-Franguiadakis}, \citenamefont {Eid}, \citenamefont {Socol}, \citenamefont {Molcrette}, \citenamefont {Guegan}, \citenamefont {Mougel}, \citenamefont {Salvetti},\ and\ \citenamefont {Montel}}]{chazot2022optical}%
  \BibitemOpen
  \bibfield  {author} {\bibinfo {author} {\bibfnamefont {L.}~\bibnamefont {Chazot-Franguiadakis}}, \bibinfo {author} {\bibfnamefont {J.}~\bibnamefont {Eid}}, \bibinfo {author} {\bibfnamefont {M.}~\bibnamefont {Socol}}, \bibinfo {author} {\bibfnamefont {B.}~\bibnamefont {Molcrette}}, \bibinfo {author} {\bibfnamefont {P.}~\bibnamefont {Guegan}}, \bibinfo {author} {\bibfnamefont {M.}~\bibnamefont {Mougel}}, \bibinfo {author} {\bibfnamefont {A.}~\bibnamefont {Salvetti}}, \ and\ \bibinfo {author} {\bibfnamefont {F.}~\bibnamefont {Montel}},\ }\bibfield  {title} {\enquote {\bibinfo {title} {Optical quantification by nanopores of viruses, extracellular vesicles, and nanoparticles},}\ }\href@noop {} {\bibfield  {journal} {\bibinfo  {journal} {Nano Letters}\ }\textbf {\bibinfo {volume} {22}},\ \bibinfo {pages} {3651--3658} (\bibinfo {year} {2022})}\BibitemShut {NoStop}%
\bibitem [{\citenamefont {Kavokine}, \citenamefont {Netz},\ and\ \citenamefont {Bocquet}(2021)}]{kavokine2021fluids}%
  \BibitemOpen
  \bibfield  {author} {\bibinfo {author} {\bibfnamefont {N.}~\bibnamefont {Kavokine}}, \bibinfo {author} {\bibfnamefont {R.~R.}\ \bibnamefont {Netz}}, \ and\ \bibinfo {author} {\bibfnamefont {L.}~\bibnamefont {Bocquet}},\ }\bibfield  {title} {\enquote {\bibinfo {title} {Fluids at the nanoscale: from continuum to subcontinuum transport},}\ }\href@noop {} {\bibfield  {journal} {\bibinfo  {journal} {Annual Review of Fluid Mechanics}\ }\textbf {\bibinfo {volume} {53}},\ \bibinfo {pages} {377--410} (\bibinfo {year} {2021})}\BibitemShut {NoStop}%
\bibitem [{\citenamefont {Verveen}\ and\ \citenamefont {DeFelice}(1974)}]{verveen1974membrane}%
  \BibitemOpen
  \bibfield  {author} {\bibinfo {author} {\bibfnamefont {A.}~\bibnamefont {Verveen}}\ and\ \bibinfo {author} {\bibfnamefont {L.}~\bibnamefont {DeFelice}},\ }\bibfield  {title} {\enquote {\bibinfo {title} {Membrane noise},}\ }\href@noop {} {\bibfield  {journal} {\bibinfo  {journal} {Progress in biophysics and molecular biology}\ }\textbf {\bibinfo {volume} {28}},\ \bibinfo {pages} {189--265} (\bibinfo {year} {1974})}\BibitemShut {NoStop}%
\bibitem [{\citenamefont {Luckey}\ and\ \citenamefont {Nikaido}(1980)}]{luckey1980specificity}%
  \BibitemOpen
  \bibfield  {author} {\bibinfo {author} {\bibfnamefont {M.}~\bibnamefont {Luckey}}\ and\ \bibinfo {author} {\bibfnamefont {H.}~\bibnamefont {Nikaido}},\ }\bibfield  {title} {\enquote {\bibinfo {title} {Specificity of diffusion channels produced by lambda phage receptor protein of escherichia coli.}}\ }\href@noop {} {\bibfield  {journal} {\bibinfo  {journal} {Proceedings of the National Academy of Sciences}\ }\textbf {\bibinfo {volume} {77}},\ \bibinfo {pages} {167--171} (\bibinfo {year} {1980})}\BibitemShut {NoStop}%
\bibitem [{\citenamefont {Benz}, \citenamefont {Schmid},\ and\ \citenamefont {Vos-Scheperkeuter}(1987)}]{benz1987mechanism}%
  \BibitemOpen
  \bibfield  {author} {\bibinfo {author} {\bibfnamefont {R.}~\bibnamefont {Benz}}, \bibinfo {author} {\bibfnamefont {A.}~\bibnamefont {Schmid}}, \ and\ \bibinfo {author} {\bibfnamefont {G.~H.}\ \bibnamefont {Vos-Scheperkeuter}},\ }\bibfield  {title} {\enquote {\bibinfo {title} {Mechanism of sugar transport through the sugar-specific lamb channel of escherichia coli outer membrane},}\ }\href@noop {} {\bibfield  {journal} {\bibinfo  {journal} {The journal of membrane biology}\ }\textbf {\bibinfo {volume} {100}},\ \bibinfo {pages} {21--29} (\bibinfo {year} {1987})}\BibitemShut {NoStop}%
\bibitem [{\citenamefont {Andersen}, \citenamefont {Jordy},\ and\ \citenamefont {Benz}(1995)}]{andersen1995evaluation}%
  \BibitemOpen
  \bibfield  {author} {\bibinfo {author} {\bibfnamefont {C.}~\bibnamefont {Andersen}}, \bibinfo {author} {\bibfnamefont {M.}~\bibnamefont {Jordy}}, \ and\ \bibinfo {author} {\bibfnamefont {R.}~\bibnamefont {Benz}},\ }\bibfield  {title} {\enquote {\bibinfo {title} {Evaluation of the rate constants of sugar transport through maltoporin (lamb) of escherichia coli from the sugar-induced current noise.}}\ }\href@noop {} {\bibfield  {journal} {\bibinfo  {journal} {Journal of General Physiology}\ }\textbf {\bibinfo {volume} {105}},\ \bibinfo {pages} {385--401} (\bibinfo {year} {1995})}\BibitemShut {NoStop}%
\bibitem [{\citenamefont {Howorka}, \citenamefont {Cheley},\ and\ \citenamefont {Bayley}(2001)}]{howorka2001sequence}%
  \BibitemOpen
  \bibfield  {author} {\bibinfo {author} {\bibfnamefont {S.}~\bibnamefont {Howorka}}, \bibinfo {author} {\bibfnamefont {S.}~\bibnamefont {Cheley}}, \ and\ \bibinfo {author} {\bibfnamefont {H.}~\bibnamefont {Bayley}},\ }\bibfield  {title} {\enquote {\bibinfo {title} {Sequence-specific detection of individual dna strands using engineered nanopores},}\ }\href@noop {} {\bibfield  {journal} {\bibinfo  {journal} {Nature biotechnology}\ }\textbf {\bibinfo {volume} {19}},\ \bibinfo {pages} {636--639} (\bibinfo {year} {2001})}\BibitemShut {NoStop}%
\bibitem [{\citenamefont {Wei}, \citenamefont {Tamp{\'e}},\ and\ \citenamefont {Rant}(2012)}]{wei2012stochastic}%
  \BibitemOpen
  \bibfield  {author} {\bibinfo {author} {\bibfnamefont {R.}~\bibnamefont {Wei}}, \bibinfo {author} {\bibfnamefont {R.}~\bibnamefont {Tamp{\'e}}}, \ and\ \bibinfo {author} {\bibfnamefont {U.}~\bibnamefont {Rant}},\ }\bibfield  {title} {\enquote {\bibinfo {title} {Stochastic sensing of proteins with receptor-modified solid-state nanopores},}\ }\href@noop {} {\bibfield  {journal} {\bibinfo  {journal} {Biophysical Journal}\ }\textbf {\bibinfo {volume} {102}},\ \bibinfo {pages} {429a} (\bibinfo {year} {2012})}\BibitemShut {NoStop}%
\bibitem [{\citenamefont {Yusko}\ \emph {et~al.}(2011)\citenamefont {Yusko}, \citenamefont {Johnson}, \citenamefont {Majd}, \citenamefont {Prangkio}, \citenamefont {Rollings}, \citenamefont {Li}, \citenamefont {Yang},\ and\ \citenamefont {Mayer}}]{yusko2011controlling}%
  \BibitemOpen
  \bibfield  {author} {\bibinfo {author} {\bibfnamefont {E.~C.}\ \bibnamefont {Yusko}}, \bibinfo {author} {\bibfnamefont {J.~M.}\ \bibnamefont {Johnson}}, \bibinfo {author} {\bibfnamefont {S.}~\bibnamefont {Majd}}, \bibinfo {author} {\bibfnamefont {P.}~\bibnamefont {Prangkio}}, \bibinfo {author} {\bibfnamefont {R.~C.}\ \bibnamefont {Rollings}}, \bibinfo {author} {\bibfnamefont {J.}~\bibnamefont {Li}}, \bibinfo {author} {\bibfnamefont {J.}~\bibnamefont {Yang}}, \ and\ \bibinfo {author} {\bibfnamefont {M.}~\bibnamefont {Mayer}},\ }\bibfield  {title} {\enquote {\bibinfo {title} {Controlling protein translocation through nanopores with bio-inspired fluid walls},}\ }\href@noop {} {\bibfield  {journal} {\bibinfo  {journal} {Nature nanotechnology}\ }\textbf {\bibinfo {volume} {6}},\ \bibinfo {pages} {253--260} (\bibinfo {year} {2011})}\BibitemShut {NoStop}%
\bibitem [{\citenamefont {Rotem}\ \emph {et~al.}(2012)\citenamefont {Rotem}, \citenamefont {Jayasinghe}, \citenamefont {Salichou},\ and\ \citenamefont {Bayley}}]{rotem2012protein}%
  \BibitemOpen
  \bibfield  {author} {\bibinfo {author} {\bibfnamefont {D.}~\bibnamefont {Rotem}}, \bibinfo {author} {\bibfnamefont {L.}~\bibnamefont {Jayasinghe}}, \bibinfo {author} {\bibfnamefont {M.}~\bibnamefont {Salichou}}, \ and\ \bibinfo {author} {\bibfnamefont {H.}~\bibnamefont {Bayley}},\ }\bibfield  {title} {\enquote {\bibinfo {title} {Protein detection by nanopores equipped with aptamers},}\ }\href@noop {} {\bibfield  {journal} {\bibinfo  {journal} {Journal of the American Chemical Society}\ }\textbf {\bibinfo {volume} {134}},\ \bibinfo {pages} {2781--2787} (\bibinfo {year} {2012})}\BibitemShut {NoStop}%
\bibitem [{\citenamefont {Wiggin}\ \emph {et~al.}(2008)\citenamefont {Wiggin}, \citenamefont {Tropini}, \citenamefont {Tabard-Cossa}, \citenamefont {Jetha},\ and\ \citenamefont {Marziali}}]{wiggin2008nonexponential}%
  \BibitemOpen
  \bibfield  {author} {\bibinfo {author} {\bibfnamefont {M.}~\bibnamefont {Wiggin}}, \bibinfo {author} {\bibfnamefont {C.}~\bibnamefont {Tropini}}, \bibinfo {author} {\bibfnamefont {V.}~\bibnamefont {Tabard-Cossa}}, \bibinfo {author} {\bibfnamefont {N.~N.}\ \bibnamefont {Jetha}}, \ and\ \bibinfo {author} {\bibfnamefont {A.}~\bibnamefont {Marziali}},\ }\bibfield  {title} {\enquote {\bibinfo {title} {Nonexponential kinetics of dna escape from $\alpha$-hemolysin nanopores},}\ }\href@noop {} {\bibfield  {journal} {\bibinfo  {journal} {Biophysical journal}\ }\textbf {\bibinfo {volume} {95}},\ \bibinfo {pages} {5317--5323} (\bibinfo {year} {2008})}\BibitemShut {NoStop}%
\bibitem [{\citenamefont {Jetha}\ \emph {et~al.}(2011)\citenamefont {Jetha}, \citenamefont {Feehan}, \citenamefont {Wiggin}, \citenamefont {Tabard-Cossa},\ and\ \citenamefont {Marziali}}]{jetha2011long}%
  \BibitemOpen
  \bibfield  {author} {\bibinfo {author} {\bibfnamefont {N.~N.}\ \bibnamefont {Jetha}}, \bibinfo {author} {\bibfnamefont {C.}~\bibnamefont {Feehan}}, \bibinfo {author} {\bibfnamefont {M.}~\bibnamefont {Wiggin}}, \bibinfo {author} {\bibfnamefont {V.}~\bibnamefont {Tabard-Cossa}}, \ and\ \bibinfo {author} {\bibfnamefont {A.}~\bibnamefont {Marziali}},\ }\bibfield  {title} {\enquote {\bibinfo {title} {Long dwell-time passage of dna through nanometer-scale pores: kinetics and sequence dependence of motion},}\ }\href@noop {} {\bibfield  {journal} {\bibinfo  {journal} {Biophysical Journal}\ }\textbf {\bibinfo {volume} {100}},\ \bibinfo {pages} {2974--2980} (\bibinfo {year} {2011})}\BibitemShut {NoStop}%
\bibitem [{\citenamefont {Singh}\ \emph {et~al.}(2011)\citenamefont {Singh}, \citenamefont {Chan}, \citenamefont {Kang},\ and\ \citenamefont {Lemay}}]{singh2011stochastic}%
  \BibitemOpen
  \bibfield  {author} {\bibinfo {author} {\bibfnamefont {P.~S.}\ \bibnamefont {Singh}}, \bibinfo {author} {\bibfnamefont {H.-S.~M.}\ \bibnamefont {Chan}}, \bibinfo {author} {\bibfnamefont {S.}~\bibnamefont {Kang}}, \ and\ \bibinfo {author} {\bibfnamefont {S.~G.}\ \bibnamefont {Lemay}},\ }\bibfield  {title} {\enquote {\bibinfo {title} {Stochastic amperometric fluctuations as a probe for dynamic adsorption in nanofluidic electrochemical systems},}\ }\href@noop {} {\bibfield  {journal} {\bibinfo  {journal} {Journal of the American Chemical Society}\ }\textbf {\bibinfo {volume} {133}},\ \bibinfo {pages} {18289--18295} (\bibinfo {year} {2011})}\BibitemShut {NoStop}%
\bibitem [{\citenamefont {Nekolla}, \citenamefont {Andersen},\ and\ \citenamefont {Benz}(1994)}]{nekolla1994noise}%
  \BibitemOpen
  \bibfield  {author} {\bibinfo {author} {\bibfnamefont {S.}~\bibnamefont {Nekolla}}, \bibinfo {author} {\bibfnamefont {C.}~\bibnamefont {Andersen}}, \ and\ \bibinfo {author} {\bibfnamefont {R.}~\bibnamefont {Benz}},\ }\bibfield  {title} {\enquote {\bibinfo {title} {Noise analysis of ion current through the open and the sugar-induced closed state of the lamb channel of escherichia coli outer membrane: evaluation of the sugar binding kinetics to the channel interior},}\ }\href@noop {} {\bibfield  {journal} {\bibinfo  {journal} {Biophysical journal}\ }\textbf {\bibinfo {volume} {66}},\ \bibinfo {pages} {1388--1397} (\bibinfo {year} {1994})}\BibitemShut {NoStop}%
\bibitem [{\citenamefont {Hoogerheide}, \citenamefont {Garaj},\ and\ \citenamefont {Golovchenko}(2009)}]{hoogerheide2009probing}%
  \BibitemOpen
  \bibfield  {author} {\bibinfo {author} {\bibfnamefont {D.~P.}\ \bibnamefont {Hoogerheide}}, \bibinfo {author} {\bibfnamefont {S.}~\bibnamefont {Garaj}}, \ and\ \bibinfo {author} {\bibfnamefont {J.~A.}\ \bibnamefont {Golovchenko}},\ }\bibfield  {title} {\enquote {\bibinfo {title} {Probing surface charge fluctuations with solid-state nanopores},}\ }\href@noop {} {\bibfield  {journal} {\bibinfo  {journal} {Physical review letters}\ }\textbf {\bibinfo {volume} {102}},\ \bibinfo {pages} {256804} (\bibinfo {year} {2009})}\BibitemShut {NoStop}%
\bibitem [{\citenamefont {Secchi}\ \emph {et~al.}(2016)\citenamefont {Secchi}, \citenamefont {Nigu{\`e}s}, \citenamefont {Jubin}, \citenamefont {Siria},\ and\ \citenamefont {Bocquet}}]{secchi2016scaling}%
  \BibitemOpen
  \bibfield  {author} {\bibinfo {author} {\bibfnamefont {E.}~\bibnamefont {Secchi}}, \bibinfo {author} {\bibfnamefont {A.}~\bibnamefont {Nigu{\`e}s}}, \bibinfo {author} {\bibfnamefont {L.}~\bibnamefont {Jubin}}, \bibinfo {author} {\bibfnamefont {A.}~\bibnamefont {Siria}}, \ and\ \bibinfo {author} {\bibfnamefont {L.}~\bibnamefont {Bocquet}},\ }\bibfield  {title} {\enquote {\bibinfo {title} {Scaling behavior for ionic transport and its fluctuations in individual carbon nanotubes},}\ }\href@noop {} {\bibfield  {journal} {\bibinfo  {journal} {Physical review letters}\ }\textbf {\bibinfo {volume} {116}},\ \bibinfo {pages} {154501} (\bibinfo {year} {2016})}\BibitemShut {NoStop}%
\bibitem [{\citenamefont {Bezrukov}\ and\ \citenamefont {Kasianowicz}(1993)}]{bezrukov1993current}%
  \BibitemOpen
  \bibfield  {author} {\bibinfo {author} {\bibfnamefont {S.~M.}\ \bibnamefont {Bezrukov}}\ and\ \bibinfo {author} {\bibfnamefont {J.~J.}\ \bibnamefont {Kasianowicz}},\ }\bibfield  {title} {\enquote {\bibinfo {title} {Current noise reveals protonation kinetics and number of ionizable sites in an open protein ion channel},}\ }\href@noop {} {\bibfield  {journal} {\bibinfo  {journal} {Physical review letters}\ }\textbf {\bibinfo {volume} {70}},\ \bibinfo {pages} {2352} (\bibinfo {year} {1993})}\BibitemShut {NoStop}%
\bibitem [{\citenamefont {Knowles}\ \emph {et~al.}(2021)\citenamefont {Knowles}, \citenamefont {Weckman}, \citenamefont {Lim}, \citenamefont {Bonthuis}, \citenamefont {Keyser},\ and\ \citenamefont {Thorneywork}}]{knowles2021current}%
  \BibitemOpen
  \bibfield  {author} {\bibinfo {author} {\bibfnamefont {S.~F.}\ \bibnamefont {Knowles}}, \bibinfo {author} {\bibfnamefont {N.~E.}\ \bibnamefont {Weckman}}, \bibinfo {author} {\bibfnamefont {V.~J.}\ \bibnamefont {Lim}}, \bibinfo {author} {\bibfnamefont {D.~J.}\ \bibnamefont {Bonthuis}}, \bibinfo {author} {\bibfnamefont {U.~F.}\ \bibnamefont {Keyser}}, \ and\ \bibinfo {author} {\bibfnamefont {A.~L.}\ \bibnamefont {Thorneywork}},\ }\bibfield  {title} {\enquote {\bibinfo {title} {Current fluctuations in nanopores reveal the polymer-wall adsorption potential},}\ }\href@noop {} {\bibfield  {journal} {\bibinfo  {journal} {Physical Review Letters}\ }\textbf {\bibinfo {volume} {127}},\ \bibinfo {pages} {137801} (\bibinfo {year} {2021})}\BibitemShut {NoStop}%
\bibitem [{\citenamefont {Queralt-Mart{\'\i}n}, \citenamefont {Perini},\ and\ \citenamefont {Alcaraz}(2021)}]{queralt2021specific}%
  \BibitemOpen
  \bibfield  {author} {\bibinfo {author} {\bibfnamefont {M.}~\bibnamefont {Queralt-Mart{\'\i}n}}, \bibinfo {author} {\bibfnamefont {D.~A.}\ \bibnamefont {Perini}}, \ and\ \bibinfo {author} {\bibfnamefont {A.}~\bibnamefont {Alcaraz}},\ }\bibfield  {title} {\enquote {\bibinfo {title} {Specific adsorption of trivalent cations in biological nanopores determines conductance dynamics and reverses ionic selectivity},}\ }\href@noop {} {\bibfield  {journal} {\bibinfo  {journal} {Physical Chemistry Chemical Physics}\ }\textbf {\bibinfo {volume} {23}},\ \bibinfo {pages} {1352--1362} (\bibinfo {year} {2021})}\BibitemShut {NoStop}%
\bibitem [{\citenamefont {Gravelle}, \citenamefont {Netz},\ and\ \citenamefont {Bocquet}(2019)}]{gravelle2019adsorption}%
  \BibitemOpen
  \bibfield  {author} {\bibinfo {author} {\bibfnamefont {S.}~\bibnamefont {Gravelle}}, \bibinfo {author} {\bibfnamefont {R.~R.}\ \bibnamefont {Netz}}, \ and\ \bibinfo {author} {\bibfnamefont {L.}~\bibnamefont {Bocquet}},\ }\bibfield  {title} {\enquote {\bibinfo {title} {Adsorption kinetics in open nanopores as a source of low-frequency noise},}\ }\href@noop {} {\bibfield  {journal} {\bibinfo  {journal} {Nano letters}\ }\textbf {\bibinfo {volume} {19}},\ \bibinfo {pages} {7265--7272} (\bibinfo {year} {2019})}\BibitemShut {NoStop}%
\bibitem [{\citenamefont {K\"atelh\"on}\ \emph {et~al.}(2014)\citenamefont {K\"atelh\"on}, \citenamefont {Krause}, \citenamefont {Mathwig}, \citenamefont {Lemay},\ and\ \citenamefont {Wolfrum}}]{katelhon2014noise}%
  \BibitemOpen
  \bibfield  {author} {\bibinfo {author} {\bibfnamefont {E.}~\bibnamefont {K\"atelh\"on}}, \bibinfo {author} {\bibfnamefont {K.~J.}\ \bibnamefont {Krause}}, \bibinfo {author} {\bibfnamefont {K.}~\bibnamefont {Mathwig}}, \bibinfo {author} {\bibfnamefont {S.~G.}\ \bibnamefont {Lemay}}, \ and\ \bibinfo {author} {\bibfnamefont {B.}~\bibnamefont {Wolfrum}},\ }\bibfield  {title} {\enquote {\bibinfo {title} {Noise phenomena caused by reversible adsorption in nanoscale electrochemical devices},}\ }\href@noop {} {\bibfield  {journal} {\bibinfo  {journal} {ACS nano}\ }\textbf {\bibinfo {volume} {8}},\ \bibinfo {pages} {4924--4930} (\bibinfo {year} {2014})}\BibitemShut {NoStop}%
\bibitem [{\citenamefont {Zevenbergen}\ \emph {et~al.}(2009)\citenamefont {Zevenbergen}, \citenamefont {Singh}, \citenamefont {Goluch}, \citenamefont {Wolfrum},\ and\ \citenamefont {Lemay}}]{zevenbergen2009electrochemical}%
  \BibitemOpen
  \bibfield  {author} {\bibinfo {author} {\bibfnamefont {M.~A.}\ \bibnamefont {Zevenbergen}}, \bibinfo {author} {\bibfnamefont {P.~S.}\ \bibnamefont {Singh}}, \bibinfo {author} {\bibfnamefont {E.~D.}\ \bibnamefont {Goluch}}, \bibinfo {author} {\bibfnamefont {B.~L.}\ \bibnamefont {Wolfrum}}, \ and\ \bibinfo {author} {\bibfnamefont {S.~G.}\ \bibnamefont {Lemay}},\ }\bibfield  {title} {\enquote {\bibinfo {title} {Electrochemical correlation spectroscopy in nanofluidic cavities},}\ }\href@noop {} {\bibfield  {journal} {\bibinfo  {journal} {Analytical chemistry}\ }\textbf {\bibinfo {volume} {81}},\ \bibinfo {pages} {8203--8212} (\bibinfo {year} {2009})}\BibitemShut {NoStop}%
\bibitem [{\citenamefont {Smeets}\ \emph {et~al.}(2008)\citenamefont {Smeets}, \citenamefont {Keyser}, \citenamefont {Dekker},\ and\ \citenamefont {Dekker}}]{smeets2008noise}%
  \BibitemOpen
  \bibfield  {author} {\bibinfo {author} {\bibfnamefont {R.~M.}\ \bibnamefont {Smeets}}, \bibinfo {author} {\bibfnamefont {U.~F.}\ \bibnamefont {Keyser}}, \bibinfo {author} {\bibfnamefont {N.~H.}\ \bibnamefont {Dekker}}, \ and\ \bibinfo {author} {\bibfnamefont {C.}~\bibnamefont {Dekker}},\ }\bibfield  {title} {\enquote {\bibinfo {title} {Noise in solid-state nanopores},}\ }\href@noop {} {\bibfield  {journal} {\bibinfo  {journal} {Proceedings of the National Academy of Sciences}\ }\textbf {\bibinfo {volume} {105}},\ \bibinfo {pages} {417--421} (\bibinfo {year} {2008})}\BibitemShut {NoStop}%
\bibitem [{\citenamefont {Smeets}, \citenamefont {Dekker},\ and\ \citenamefont {Dekker}(2009)}]{smeets2009low}%
  \BibitemOpen
  \bibfield  {author} {\bibinfo {author} {\bibfnamefont {R.}~\bibnamefont {Smeets}}, \bibinfo {author} {\bibfnamefont {N.}~\bibnamefont {Dekker}}, \ and\ \bibinfo {author} {\bibfnamefont {C.}~\bibnamefont {Dekker}},\ }\bibfield  {title} {\enquote {\bibinfo {title} {Low-frequency noise in solid-state nanopores},}\ }\href@noop {} {\bibfield  {journal} {\bibinfo  {journal} {Nanotechnology}\ }\textbf {\bibinfo {volume} {20}},\ \bibinfo {pages} {095501} (\bibinfo {year} {2009})}\BibitemShut {NoStop}%
\bibitem [{\citenamefont {Knowles}, \citenamefont {Keyser},\ and\ \citenamefont {Thorneywork}(2019)}]{knowles2019noise}%
  \BibitemOpen
  \bibfield  {author} {\bibinfo {author} {\bibfnamefont {S.}~\bibnamefont {Knowles}}, \bibinfo {author} {\bibfnamefont {U.}~\bibnamefont {Keyser}}, \ and\ \bibinfo {author} {\bibfnamefont {A.}~\bibnamefont {Thorneywork}},\ }\bibfield  {title} {\enquote {\bibinfo {title} {Noise properties of rectifying and non-rectifying nanopores},}\ }\href@noop {} {\bibfield  {journal} {\bibinfo  {journal} {Nanotechnology}\ }\textbf {\bibinfo {volume} {31}},\ \bibinfo {pages} {10LT01} (\bibinfo {year} {2019})}\BibitemShut {NoStop}%
\bibitem [{\citenamefont {Fuli{\'n}ski}\ \emph {et~al.}(1998)\citenamefont {Fuli{\'n}ski}, \citenamefont {Grzywna}, \citenamefont {Mellor}, \citenamefont {Siwy},\ and\ \citenamefont {Usherwood}}]{fulinski1998non}%
  \BibitemOpen
  \bibfield  {author} {\bibinfo {author} {\bibfnamefont {A.}~\bibnamefont {Fuli{\'n}ski}}, \bibinfo {author} {\bibfnamefont {Z.}~\bibnamefont {Grzywna}}, \bibinfo {author} {\bibfnamefont {I.}~\bibnamefont {Mellor}}, \bibinfo {author} {\bibfnamefont {Z.}~\bibnamefont {Siwy}}, \ and\ \bibinfo {author} {\bibfnamefont {P.}~\bibnamefont {Usherwood}},\ }\bibfield  {title} {\enquote {\bibinfo {title} {Non-markovian character of ionic current fluctuations in membrane channels},}\ }\href@noop {} {\bibfield  {journal} {\bibinfo  {journal} {Physical Review E}\ }\textbf {\bibinfo {volume} {58}},\ \bibinfo {pages} {919} (\bibinfo {year} {1998})}\BibitemShut {NoStop}%
\bibitem [{\citenamefont {Fragasso}, \citenamefont {Pud},\ and\ \citenamefont {Dekker}(2019)}]{fragasso20191}%
  \BibitemOpen
  \bibfield  {author} {\bibinfo {author} {\bibfnamefont {A.}~\bibnamefont {Fragasso}}, \bibinfo {author} {\bibfnamefont {S.}~\bibnamefont {Pud}}, \ and\ \bibinfo {author} {\bibfnamefont {C.}~\bibnamefont {Dekker}},\ }\bibfield  {title} {\enquote {\bibinfo {title} {1/f noise in solid-state nanopores is governed by access and surface regions},}\ }\href@noop {} {\bibfield  {journal} {\bibinfo  {journal} {Nanotechnology}\ }\textbf {\bibinfo {volume} {30}},\ \bibinfo {pages} {395202} (\bibinfo {year} {2019})}\BibitemShut {NoStop}%
\bibitem [{\citenamefont {Powell}\ \emph {et~al.}(2009)\citenamefont {Powell}, \citenamefont {Vlassiouk}, \citenamefont {Martens},\ and\ \citenamefont {Siwy}}]{powell2009nonequilibrium}%
  \BibitemOpen
  \bibfield  {author} {\bibinfo {author} {\bibfnamefont {M.~R.}\ \bibnamefont {Powell}}, \bibinfo {author} {\bibfnamefont {I.}~\bibnamefont {Vlassiouk}}, \bibinfo {author} {\bibfnamefont {C.}~\bibnamefont {Martens}}, \ and\ \bibinfo {author} {\bibfnamefont {Z.~S.}\ \bibnamefont {Siwy}},\ }\bibfield  {title} {\enquote {\bibinfo {title} {Nonequilibrium 1/f noise in rectifying nanopores},}\ }\href@noop {} {\bibfield  {journal} {\bibinfo  {journal} {Physical review letters}\ }\textbf {\bibinfo {volume} {103}},\ \bibinfo {pages} {248104} (\bibinfo {year} {2009})}\BibitemShut {NoStop}%
\bibitem [{\citenamefont {Wohnsland}\ and\ \citenamefont {Benz}(1997)}]{wohnsland19971}%
  \BibitemOpen
  \bibfield  {author} {\bibinfo {author} {\bibfnamefont {F.}~\bibnamefont {Wohnsland}}\ and\ \bibinfo {author} {\bibfnamefont {R.}~\bibnamefont {Benz}},\ }\bibfield  {title} {\enquote {\bibinfo {title} {1/f-noise of open bacterial porin channels},}\ }\href@noop {} {\bibfield  {journal} {\bibinfo  {journal} {The Journal of membrane biology}\ }\textbf {\bibinfo {volume} {158}},\ \bibinfo {pages} {77--85} (\bibinfo {year} {1997})}\BibitemShut {NoStop}%
\bibitem [{\citenamefont {Siwy}\ and\ \citenamefont {Fuli{\'n}ski}(2002)}]{siwy2002origin}%
  \BibitemOpen
  \bibfield  {author} {\bibinfo {author} {\bibfnamefont {Z.}~\bibnamefont {Siwy}}\ and\ \bibinfo {author} {\bibfnamefont {A.}~\bibnamefont {Fuli{\'n}ski}},\ }\bibfield  {title} {\enquote {\bibinfo {title} {Origin of 1/f $\alpha$ noise in membrane channel currents},}\ }\href@noop {} {\bibfield  {journal} {\bibinfo  {journal} {Physical Review Letters}\ }\textbf {\bibinfo {volume} {89}},\ \bibinfo {pages} {158101} (\bibinfo {year} {2002})}\BibitemShut {NoStop}%
\bibitem [{\citenamefont {Bezrukov}\ \emph {et~al.}(2000)\citenamefont {Bezrukov}, \citenamefont {Berezhkovskii}, \citenamefont {Pustovoit},\ and\ \citenamefont {Szabo}}]{bezrukov2000particle}%
  \BibitemOpen
  \bibfield  {author} {\bibinfo {author} {\bibfnamefont {S.~M.}\ \bibnamefont {Bezrukov}}, \bibinfo {author} {\bibfnamefont {A.~M.}\ \bibnamefont {Berezhkovskii}}, \bibinfo {author} {\bibfnamefont {M.~A.}\ \bibnamefont {Pustovoit}}, \ and\ \bibinfo {author} {\bibfnamefont {A.}~\bibnamefont {Szabo}},\ }\bibfield  {title} {\enquote {\bibinfo {title} {Particle number fluctuations in a membrane channel},}\ }\href@noop {} {\bibfield  {journal} {\bibinfo  {journal} {The Journal of Chemical Physics}\ }\textbf {\bibinfo {volume} {113}},\ \bibinfo {pages} {8206--8211} (\bibinfo {year} {2000})}\BibitemShut {NoStop}%
\bibitem [{\citenamefont {Berezhkovskii}, \citenamefont {Pustovoit},\ and\ \citenamefont {Bezrukov}(2002)}]{berezhkovskii2002effect}%
  \BibitemOpen
  \bibfield  {author} {\bibinfo {author} {\bibfnamefont {A.~M.}\ \bibnamefont {Berezhkovskii}}, \bibinfo {author} {\bibfnamefont {M.~A.}\ \bibnamefont {Pustovoit}}, \ and\ \bibinfo {author} {\bibfnamefont {S.~M.}\ \bibnamefont {Bezrukov}},\ }\bibfield  {title} {\enquote {\bibinfo {title} {Effect of binding on particle number fluctuations in a membrane channel},}\ }\href@noop {} {\bibfield  {journal} {\bibinfo  {journal} {The Journal of chemical physics}\ }\textbf {\bibinfo {volume} {116}},\ \bibinfo {pages} {6216--6220} (\bibinfo {year} {2002})}\BibitemShut {NoStop}%
\bibitem [{\citenamefont {Marbach}(2021)}]{marbach2021intrinsic}%
  \BibitemOpen
  \bibfield  {author} {\bibinfo {author} {\bibfnamefont {S.}~\bibnamefont {Marbach}},\ }\bibfield  {title} {\enquote {\bibinfo {title} {Intrinsic fractional noise in nanopores: The effect of reservoirs},}\ }\href@noop {} {\bibfield  {journal} {\bibinfo  {journal} {The Journal of Chemical Physics}\ }\textbf {\bibinfo {volume} {154}} (\bibinfo {year} {2021})}\BibitemShut {NoStop}%
\bibitem [{\citenamefont {Nestorovich}\ \emph {et~al.}(2002)\citenamefont {Nestorovich}, \citenamefont {Danelon}, \citenamefont {Winterhalter},\ and\ \citenamefont {Bezrukov}}]{nestorovich2002designed}%
  \BibitemOpen
  \bibfield  {author} {\bibinfo {author} {\bibfnamefont {E.~M.}\ \bibnamefont {Nestorovich}}, \bibinfo {author} {\bibfnamefont {C.}~\bibnamefont {Danelon}}, \bibinfo {author} {\bibfnamefont {M.}~\bibnamefont {Winterhalter}}, \ and\ \bibinfo {author} {\bibfnamefont {S.~M.}\ \bibnamefont {Bezrukov}},\ }\bibfield  {title} {\enquote {\bibinfo {title} {Designed to penetrate: time-resolved interaction of single antibiotic molecules with bacterial pores},}\ }\href@noop {} {\bibfield  {journal} {\bibinfo  {journal} {Proceedings of the National Academy of Sciences}\ }\textbf {\bibinfo {volume} {99}},\ \bibinfo {pages} {9789--9794} (\bibinfo {year} {2002})}\BibitemShut {NoStop}%
\bibitem [{\citenamefont {Zorkot}, \citenamefont {Golestanian},\ and\ \citenamefont {Bonthuis}(2016{\natexlab{a}})}]{zorkot2016current}%
  \BibitemOpen
  \bibfield  {author} {\bibinfo {author} {\bibfnamefont {M.}~\bibnamefont {Zorkot}}, \bibinfo {author} {\bibfnamefont {R.}~\bibnamefont {Golestanian}}, \ and\ \bibinfo {author} {\bibfnamefont {D.~J.}\ \bibnamefont {Bonthuis}},\ }\bibfield  {title} {\enquote {\bibinfo {title} {Current fluctuations in nanopores: The effects of electrostatic and hydrodynamic interactions},}\ }\href@noop {} {\bibfield  {journal} {\bibinfo  {journal} {The European Physical Journal Special Topics}\ }\textbf {\bibinfo {volume} {225}},\ \bibinfo {pages} {1583--1594} (\bibinfo {year} {2016}{\natexlab{a}})}\BibitemShut {NoStop}%
\bibitem [{\citenamefont {Zorkot}, \citenamefont {Golestanian},\ and\ \citenamefont {Bonthuis}(2016{\natexlab{b}})}]{zorkot2016power}%
  \BibitemOpen
  \bibfield  {author} {\bibinfo {author} {\bibfnamefont {M.}~\bibnamefont {Zorkot}}, \bibinfo {author} {\bibfnamefont {R.}~\bibnamefont {Golestanian}}, \ and\ \bibinfo {author} {\bibfnamefont {D.~J.}\ \bibnamefont {Bonthuis}},\ }\bibfield  {title} {\enquote {\bibinfo {title} {The power spectrum of ionic nanopore currents: the role of ion correlations},}\ }\href@noop {} {\bibfield  {journal} {\bibinfo  {journal} {Nano letters}\ }\textbf {\bibinfo {volume} {16}},\ \bibinfo {pages} {2205--2212} (\bibinfo {year} {2016}{\natexlab{b}})}\BibitemShut {NoStop}%
\bibitem [{\citenamefont {Knowles}, \citenamefont {Mackay},\ and\ \citenamefont {Thorneywork}(2024)}]{knowles2024interpreting}%
  \BibitemOpen
  \bibfield  {author} {\bibinfo {author} {\bibfnamefont {S.~F.}\ \bibnamefont {Knowles}}, \bibinfo {author} {\bibfnamefont {E.~K.}\ \bibnamefont {Mackay}}, \ and\ \bibinfo {author} {\bibfnamefont {A.~L.}\ \bibnamefont {Thorneywork}},\ }\bibfield  {title} {\enquote {\bibinfo {title} {Interpreting the power spectral density of a fluctuating colloidal current},}\ }\href@noop {} {\bibfield  {journal} {\bibinfo  {journal} {The Journal of Chemical Physics}\ }\textbf {\bibinfo {volume} {161}} (\bibinfo {year} {2024})}\BibitemShut {NoStop}%
\bibitem [{\citenamefont {Robin}\ \emph {et~al.}(2023{\natexlab{a}})\citenamefont {Robin}, \citenamefont {Liz{\'e}e}, \citenamefont {Yang}, \citenamefont {Emmerich}, \citenamefont {Siria},\ and\ \citenamefont {Bocquet}}]{robin2023disentangling}%
  \BibitemOpen
  \bibfield  {author} {\bibinfo {author} {\bibfnamefont {P.}~\bibnamefont {Robin}}, \bibinfo {author} {\bibfnamefont {M.}~\bibnamefont {Liz{\'e}e}}, \bibinfo {author} {\bibfnamefont {Q.}~\bibnamefont {Yang}}, \bibinfo {author} {\bibfnamefont {T.}~\bibnamefont {Emmerich}}, \bibinfo {author} {\bibfnamefont {A.}~\bibnamefont {Siria}}, \ and\ \bibinfo {author} {\bibfnamefont {L.}~\bibnamefont {Bocquet}},\ }\bibfield  {title} {\enquote {\bibinfo {title} {Disentangling 1/f noise from confined ion dynamics},}\ }\href@noop {} {\bibfield  {journal} {\bibinfo  {journal} {Faraday Discussions}\ }\textbf {\bibinfo {volume} {246}},\ \bibinfo {pages} {556--575} (\bibinfo {year} {2023}{\natexlab{a}})}\BibitemShut {NoStop}%
\bibitem [{\citenamefont {Gadgil}\ \emph {et~al.}(2004)\citenamefont {Gadgil}, \citenamefont {Yeckel}, \citenamefont {Derby},\ and\ \citenamefont {Hu}}]{gadgil2004diffusion}%
  \BibitemOpen
  \bibfield  {author} {\bibinfo {author} {\bibfnamefont {C.}~\bibnamefont {Gadgil}}, \bibinfo {author} {\bibfnamefont {A.}~\bibnamefont {Yeckel}}, \bibinfo {author} {\bibfnamefont {J.~J.}\ \bibnamefont {Derby}}, \ and\ \bibinfo {author} {\bibfnamefont {W.-S.}\ \bibnamefont {Hu}},\ }\bibfield  {title} {\enquote {\bibinfo {title} {A diffusion--reaction model for dna microarray assays},}\ }\href@noop {} {\bibfield  {journal} {\bibinfo  {journal} {Journal of biotechnology}\ }\textbf {\bibinfo {volume} {114}},\ \bibinfo {pages} {31--45} (\bibinfo {year} {2004})}\BibitemShut {NoStop}%
\bibitem [{\citenamefont {Bendat}\ and\ \citenamefont {Piersol}(2010)}]{alma991022195870107026}%
  \BibitemOpen
  \bibfield  {author} {\bibinfo {author} {\bibfnamefont {J.~S.}\ \bibnamefont {Bendat}}\ and\ \bibinfo {author} {\bibfnamefont {A.~G.}\ \bibnamefont {Piersol}},\ }\href@noop {} {{\emph {\bibinfo {title} {Random data : analysis and measurement procedures}}}},\ \bibinfo {edition} {4th}\ ed.,\ Wiley series in probability and statistics\ (\bibinfo  {publisher} {Wiley},\ \bibinfo {address} {Hoboken, N.J},\ \bibinfo {year} {2010})\BibitemShut {NoStop}%
\bibitem [{\citenamefont {Kullman}, \citenamefont {Winterhalter},\ and\ \citenamefont {Bezrukov}(2002)}]{kullman2002transport}%
  \BibitemOpen
  \bibfield  {author} {\bibinfo {author} {\bibfnamefont {L.}~\bibnamefont {Kullman}}, \bibinfo {author} {\bibfnamefont {M.}~\bibnamefont {Winterhalter}}, \ and\ \bibinfo {author} {\bibfnamefont {S.~M.}\ \bibnamefont {Bezrukov}},\ }\bibfield  {title} {\enquote {\bibinfo {title} {Transport of maltodextrins through maltoporin: a single-channel study},}\ }\href@noop {} {\bibfield  {journal} {\bibinfo  {journal} {Biophysical journal}\ }\textbf {\bibinfo {volume} {82}},\ \bibinfo {pages} {803--812} (\bibinfo {year} {2002})}\BibitemShut {NoStop}%
\bibitem [{\citenamefont {Gu}, \citenamefont {Cheley},\ and\ \citenamefont {Bayley}(2001)}]{gu2001prolonged}%
  \BibitemOpen
  \bibfield  {author} {\bibinfo {author} {\bibfnamefont {L.-Q.}\ \bibnamefont {Gu}}, \bibinfo {author} {\bibfnamefont {S.}~\bibnamefont {Cheley}}, \ and\ \bibinfo {author} {\bibfnamefont {H.}~\bibnamefont {Bayley}},\ }\bibfield  {title} {\enquote {\bibinfo {title} {Prolonged residence time of a noncovalent molecular adapter, $\beta$-cyclodextrin, within the lumen of mutant $\alpha$-hemolysin pores},}\ }\href@noop {} {\bibfield  {journal} {\bibinfo  {journal} {The Journal of general physiology}\ }\textbf {\bibinfo {volume} {118}},\ \bibinfo {pages} {481--494} (\bibinfo {year} {2001})}\BibitemShut {NoStop}%
\bibitem [{\citenamefont {Zhao}\ \emph {et~al.}(2016)\citenamefont {Zhao}, \citenamefont {Gaur}, \citenamefont {Retterer}, \citenamefont {Laibinis},\ and\ \citenamefont {Weiss}}]{zhao2016flow}%
  \BibitemOpen
  \bibfield  {author} {\bibinfo {author} {\bibfnamefont {Y.}~\bibnamefont {Zhao}}, \bibinfo {author} {\bibfnamefont {G.}~\bibnamefont {Gaur}}, \bibinfo {author} {\bibfnamefont {S.~T.}\ \bibnamefont {Retterer}}, \bibinfo {author} {\bibfnamefont {P.~E.}\ \bibnamefont {Laibinis}}, \ and\ \bibinfo {author} {\bibfnamefont {S.~M.}\ \bibnamefont {Weiss}},\ }\bibfield  {title} {\enquote {\bibinfo {title} {Flow-through porous silicon membranes for real-time label-free biosensing},}\ }\href@noop {} {\bibfield  {journal} {\bibinfo  {journal} {Analytical chemistry}\ }\textbf {\bibinfo {volume} {88}},\ \bibinfo {pages} {10940--10948} (\bibinfo {year} {2016})}\BibitemShut {NoStop}%
\bibitem [{\citenamefont {Hansen}\ \emph {et~al.}(2009)\citenamefont {Hansen}, \citenamefont {Krishna}, \citenamefont {Van~Baten}, \citenamefont {Bell},\ and\ \citenamefont {Keil}}]{hansen2009analysis}%
  \BibitemOpen
  \bibfield  {author} {\bibinfo {author} {\bibfnamefont {N.}~\bibnamefont {Hansen}}, \bibinfo {author} {\bibfnamefont {R.}~\bibnamefont {Krishna}}, \bibinfo {author} {\bibfnamefont {J.}~\bibnamefont {Van~Baten}}, \bibinfo {author} {\bibfnamefont {A.~T.}\ \bibnamefont {Bell}}, \ and\ \bibinfo {author} {\bibfnamefont {F.}~\bibnamefont {Keil}},\ }\bibfield  {title} {\enquote {\bibinfo {title} {Analysis of diffusion limitation in the alkylation of benzene over h-zsm-5 by combining quantum chemical calculations, molecular simulations, and a continuum approach},}\ }\href@noop {} {\bibfield  {journal} {\bibinfo  {journal} {The Journal of Physical Chemistry C}\ }\textbf {\bibinfo {volume} {113}},\ \bibinfo {pages} {235--246} (\bibinfo {year} {2009})}\BibitemShut {NoStop}%
\bibitem [{\citenamefont {Lax}\ and\ \citenamefont {Mengert}(1960)}]{LAX1960248}%
  \BibitemOpen
  \bibfield  {author} {\bibinfo {author} {\bibfnamefont {M.}~\bibnamefont {Lax}}\ and\ \bibinfo {author} {\bibfnamefont {P.}~\bibnamefont {Mengert}},\ }\bibfield  {title} {\enquote {\bibinfo {title} {Influence of trapping, diffusion and recombination on carrier concentration fluctuations},}\ }\href {\doibase https://doi.org/10.1016/0022-3697(60)90237-7} {\bibfield  {journal} {\bibinfo  {journal} {Journal of Physics and Chemistry of Solids}\ }\textbf {\bibinfo {volume} {14}},\ \bibinfo {pages} {248--267} (\bibinfo {year} {1960})}\BibitemShut {NoStop}%
\bibitem [{\citenamefont {Minh}, \citenamefont {Rotenberg},\ and\ \citenamefont {Marbach}(2023)}]{minh2023ionic}%
  \BibitemOpen
  \bibfield  {author} {\bibinfo {author} {\bibfnamefont {T.~H.~N.}\ \bibnamefont {Minh}}, \bibinfo {author} {\bibfnamefont {B.}~\bibnamefont {Rotenberg}}, \ and\ \bibinfo {author} {\bibfnamefont {S.}~\bibnamefont {Marbach}},\ }\bibfield  {title} {\enquote {\bibinfo {title} {Ionic fluctuations in finite volumes: fractional noise and hyperuniformity},}\ }\href@noop {} {\bibfield  {journal} {\bibinfo  {journal} {Faraday discussions}\ }\textbf {\bibinfo {volume} {246}},\ \bibinfo {pages} {225--250} (\bibinfo {year} {2023})}\BibitemShut {NoStop}%
\bibitem [{\citenamefont {MacFarlane}(1950)}]{MacFarlane1950807}%
  \BibitemOpen
  \bibfield  {author} {\bibinfo {author} {\bibfnamefont {G.}~\bibnamefont {MacFarlane}},\ }\bibfield  {title} {\enquote {\bibinfo {title} {A theory of contact noise in semiconductors},}\ }\href {\doibase 10.1088/0370-1301/63/10/308} {\bibfield  {journal} {\bibinfo  {journal} {Proceedings of the Physical Society. Section B}\ }\textbf {\bibinfo {volume} {63}},\ \bibinfo {pages} {807 – 814} (\bibinfo {year} {1950})},\ \bibinfo {note} {cited by: 35}\BibitemShut {NoStop}%
\bibitem [{\citenamefont {Burgess}(1953)}]{Burgess1953334}%
  \BibitemOpen
  \bibfield  {author} {\bibinfo {author} {\bibfnamefont {R.}~\bibnamefont {Burgess}},\ }\bibfield  {title} {\enquote {\bibinfo {title} {Contact noise in semiconductors},}\ }\href {\doibase 10.1088/0370-1301/66/4/112} {\bibfield  {journal} {\bibinfo  {journal} {Proceedings of the Physical Society. Section B}\ }\textbf {\bibinfo {volume} {66}},\ \bibinfo {pages} {334 – 335} (\bibinfo {year} {1953})},\ \bibinfo {note} {cited by: 21}\BibitemShut {NoStop}%
\bibitem [{\citenamefont {{Van Vliet}}\ and\ \citenamefont {{Van der Ziel}}(1958)}]{VANVLIET1958415}%
  \BibitemOpen
  \bibfield  {author} {\bibinfo {author} {\bibfnamefont {K.}~\bibnamefont {{Van Vliet}}}\ and\ \bibinfo {author} {\bibfnamefont {A.}~\bibnamefont {{Van der Ziel}}},\ }\bibfield  {title} {\enquote {\bibinfo {title} {On the noise generated by diffusion mechanisms},}\ }\href {\doibase https://doi.org/10.1016/S0031-8914(58)95745-8} {\bibfield  {journal} {\bibinfo  {journal} {Physica}\ }\textbf {\bibinfo {volume} {24}},\ \bibinfo {pages} {415--421} (\bibinfo {year} {1958})}\BibitemShut {NoStop}%
\bibitem [{\citenamefont {Voss}\ and\ \citenamefont {Clarke}(1976)}]{vossflicker1976}%
  \BibitemOpen
  \bibfield  {author} {\bibinfo {author} {\bibfnamefont {R.~F.}\ \bibnamefont {Voss}}\ and\ \bibinfo {author} {\bibfnamefont {J.}~\bibnamefont {Clarke}},\ }\bibfield  {title} {\enquote {\bibinfo {title} {Flicker ($\frac{1}{f}$) noise: Equilibrium temperature and resistance fluctuations},}\ }\href {\doibase 10.1103/PhysRevB.13.556} {\bibfield  {journal} {\bibinfo  {journal} {Phys. Rev. B}\ }\textbf {\bibinfo {volume} {13}},\ \bibinfo {pages} {556--573} (\bibinfo {year} {1976})}\BibitemShut {NoStop}%
\bibitem [{\citenamefont {Machlup}(1954)}]{machlup1954noise}%
  \BibitemOpen
  \bibfield  {author} {\bibinfo {author} {\bibfnamefont {S.}~\bibnamefont {Machlup}},\ }\bibfield  {title} {\enquote {\bibinfo {title} {Noise in semiconductors: spectrum of a two-parameter random signal},}\ }\href@noop {} {\bibfield  {journal} {\bibinfo  {journal} {Journal of Applied Physics}\ }\textbf {\bibinfo {volume} {25}},\ \bibinfo {pages} {341--343} (\bibinfo {year} {1954})}\BibitemShut {NoStop}%
\bibitem [{\citenamefont {Emmerich}\ \emph {et~al.}(2024)\citenamefont {Emmerich}, \citenamefont {Teng}, \citenamefont {Ronceray}, \citenamefont {Lopriore}, \citenamefont {Chiesa}, \citenamefont {Chernev}, \citenamefont {Artemov}, \citenamefont {Di~Ventra}, \citenamefont {Kis},\ and\ \citenamefont {Radenovic}}]{emmerich2024nanofluidic}%
  \BibitemOpen
  \bibfield  {author} {\bibinfo {author} {\bibfnamefont {T.}~\bibnamefont {Emmerich}}, \bibinfo {author} {\bibfnamefont {Y.}~\bibnamefont {Teng}}, \bibinfo {author} {\bibfnamefont {N.}~\bibnamefont {Ronceray}}, \bibinfo {author} {\bibfnamefont {E.}~\bibnamefont {Lopriore}}, \bibinfo {author} {\bibfnamefont {R.}~\bibnamefont {Chiesa}}, \bibinfo {author} {\bibfnamefont {A.}~\bibnamefont {Chernev}}, \bibinfo {author} {\bibfnamefont {V.}~\bibnamefont {Artemov}}, \bibinfo {author} {\bibfnamefont {M.}~\bibnamefont {Di~Ventra}}, \bibinfo {author} {\bibfnamefont {A.}~\bibnamefont {Kis}}, \ and\ \bibinfo {author} {\bibfnamefont {A.}~\bibnamefont {Radenovic}},\ }\bibfield  {title} {\enquote {\bibinfo {title} {Nanofluidic logic with mechano--ionic memristive switches},}\ }\href@noop {} {\bibfield  {journal} {\bibinfo  {journal} {Nature Electronics}\ }\textbf {\bibinfo {volume} {7}},\ \bibinfo {pages} {271--278} (\bibinfo {year} {2024})}\BibitemShut {NoStop}%
\bibitem [{\citenamefont {Robin}\ \emph {et~al.}(2023{\natexlab{b}})\citenamefont {Robin}, \citenamefont {Emmerich}, \citenamefont {Ismail}, \citenamefont {Nigu{\`e}s}, \citenamefont {You}, \citenamefont {Nam}, \citenamefont {Keerthi}, \citenamefont {Siria}, \citenamefont {Geim}, \citenamefont {Radha} \emph {et~al.}}]{robin2023long}%
  \BibitemOpen
  \bibfield  {author} {\bibinfo {author} {\bibfnamefont {P.}~\bibnamefont {Robin}}, \bibinfo {author} {\bibfnamefont {T.}~\bibnamefont {Emmerich}}, \bibinfo {author} {\bibfnamefont {A.}~\bibnamefont {Ismail}}, \bibinfo {author} {\bibfnamefont {A.}~\bibnamefont {Nigu{\`e}s}}, \bibinfo {author} {\bibfnamefont {Y.}~\bibnamefont {You}}, \bibinfo {author} {\bibfnamefont {G.-H.}\ \bibnamefont {Nam}}, \bibinfo {author} {\bibfnamefont {A.}~\bibnamefont {Keerthi}}, \bibinfo {author} {\bibfnamefont {A.}~\bibnamefont {Siria}}, \bibinfo {author} {\bibfnamefont {A.}~\bibnamefont {Geim}}, \bibinfo {author} {\bibfnamefont {B.}~\bibnamefont {Radha}},  \emph {et~al.},\ }\bibfield  {title} {\enquote {\bibinfo {title} {Long-term memory and synapse-like dynamics in two-dimensional nanofluidic channels},}\ }\href@noop {} {\bibfield  {journal} {\bibinfo  {journal} {Science}\ }\textbf {\bibinfo {volume} {379}},\ \bibinfo {pages} {161--167} (\bibinfo {year} {2023}{\natexlab{b}})}\BibitemShut {NoStop}%
\bibitem [{\citenamefont {Kamsma}\ \emph {et~al.}(2024)\citenamefont {Kamsma}, \citenamefont {Kim}, \citenamefont {Kim}, \citenamefont {Boon}, \citenamefont {Spitoni}, \citenamefont {Park},\ and\ \citenamefont {van Roij}}]{kamsma2024brain}%
  \BibitemOpen
  \bibfield  {author} {\bibinfo {author} {\bibfnamefont {T.~M.}\ \bibnamefont {Kamsma}}, \bibinfo {author} {\bibfnamefont {J.}~\bibnamefont {Kim}}, \bibinfo {author} {\bibfnamefont {K.}~\bibnamefont {Kim}}, \bibinfo {author} {\bibfnamefont {W.~Q.}\ \bibnamefont {Boon}}, \bibinfo {author} {\bibfnamefont {C.}~\bibnamefont {Spitoni}}, \bibinfo {author} {\bibfnamefont {J.}~\bibnamefont {Park}}, \ and\ \bibinfo {author} {\bibfnamefont {R.}~\bibnamefont {van Roij}},\ }\bibfield  {title} {\enquote {\bibinfo {title} {Brain-inspired computing with fluidic iontronic nanochannels},}\ }\href@noop {} {\bibfield  {journal} {\bibinfo  {journal} {Proceedings of the National Academy of Sciences}\ }\textbf {\bibinfo {volume} {121}},\ \bibinfo {pages} {e2320242121} (\bibinfo {year} {2024})}\BibitemShut {NoStop}%
\bibitem [{\citenamefont {Scala}, \citenamefont {Voigtmann},\ and\ \citenamefont {De~Michele}(2007)}]{scala2007event}%
  \BibitemOpen
  \bibfield  {author} {\bibinfo {author} {\bibfnamefont {A.}~\bibnamefont {Scala}}, \bibinfo {author} {\bibfnamefont {T.}~\bibnamefont {Voigtmann}}, \ and\ \bibinfo {author} {\bibfnamefont {C.}~\bibnamefont {De~Michele}},\ }\bibfield  {title} {\enquote {\bibinfo {title} {Event-driven brownian dynamics for hard spheres},}\ }\href@noop {} {\bibfield  {journal} {\bibinfo  {journal} {The Journal of chemical physics}\ }\textbf {\bibinfo {volume} {126}} (\bibinfo {year} {2007})}\BibitemShut {NoStop}%
\bibitem [{\citenamefont {Welch}(2003)}]{welch2003use}%
  \BibitemOpen
  \bibfield  {author} {\bibinfo {author} {\bibfnamefont {P.}~\bibnamefont {Welch}},\ }\bibfield  {title} {\enquote {\bibinfo {title} {The use of fast fourier transform for the estimation of power spectra: A method based on time averaging over short, modified periodograms},}\ }\href@noop {} {\bibfield  {journal} {\bibinfo  {journal} {IEEE Transactions on audio and electroacoustics}\ }\textbf {\bibinfo {volume} {15}},\ \bibinfo {pages} {70--73} (\bibinfo {year} {2003})}\BibitemShut {NoStop}%
\bibitem [{\citenamefont {Ngai}\ and\ \citenamefont {Liu}(1981)}]{ngai1981dispersive}%
  \BibitemOpen
  \bibfield  {author} {\bibinfo {author} {\bibfnamefont {K.}~\bibnamefont {Ngai}}\ and\ \bibinfo {author} {\bibfnamefont {F.-S.}\ \bibnamefont {Liu}},\ }\bibfield  {title} {\enquote {\bibinfo {title} {Dispersive diffusion transport and noise, time-dependent diffusion coefficient, generalized einstein-nernst relation, and dispersive diffusion-controlled unimolecular and bimolecular reactions},}\ }\href@noop {} {\bibfield  {journal} {\bibinfo  {journal} {Physical Review B}\ }\textbf {\bibinfo {volume} {24}},\ \bibinfo {pages} {1049} (\bibinfo {year} {1981})}\BibitemShut {NoStop}%
\bibitem [{\citenamefont {Lax}(1960)}]{lax1960fluctuations}%
  \BibitemOpen
  \bibfield  {author} {\bibinfo {author} {\bibfnamefont {M.}~\bibnamefont {Lax}},\ }\bibfield  {title} {\enquote {\bibinfo {title} {Fluctuations from the nonequilibrium steady state},}\ }\href@noop {} {\bibfield  {journal} {\bibinfo  {journal} {Reviews of modern physics}\ }\textbf {\bibinfo {volume} {32}},\ \bibinfo {pages} {25} (\bibinfo {year} {1960})}\BibitemShut {NoStop}%
\bibitem [{\citenamefont {Hu}, \citenamefont {Gao},\ and\ \citenamefont {Li}(2007)}]{hu2007modeling}%
  \BibitemOpen
  \bibfield  {author} {\bibinfo {author} {\bibfnamefont {G.}~\bibnamefont {Hu}}, \bibinfo {author} {\bibfnamefont {Y.}~\bibnamefont {Gao}}, \ and\ \bibinfo {author} {\bibfnamefont {D.}~\bibnamefont {Li}},\ }\bibfield  {title} {\enquote {\bibinfo {title} {Modeling micropatterned antigen--antibody binding kinetics in a microfluidic chip},}\ }\href@noop {} {\bibfield  {journal} {\bibinfo  {journal} {Biosensors and Bioelectronics}\ }\textbf {\bibinfo {volume} {22}},\ \bibinfo {pages} {1403--1409} (\bibinfo {year} {2007})}\BibitemShut {NoStop}%
\end{thebibliography}
\end{document}